\documentclass{article}
\usepackage{graphicx}  
\usepackage{amsmath}   
\usepackage[compress]{cite}
\usepackage{amssymb}   
\usepackage{bm} 
\usepackage{dcolumn}
\usepackage{color}
\usepackage{mathrsfs}
\usepackage{amsfonts}
\usepackage{varioref}
\usepackage[utf8]{inputenc}
\usepackage[caption=false]{subfig}
\RequirePackage[colorlinks,citecolor=blue,urlcolor=magenta,linkcolor=blue]{hyperref}
\def\EH{Einstein-Hilbert }

\def\gr{general relativity}

\def\GPB{Gravity Probe B}
\def\FS{Frenet-Serret }
\def\KR{Kalb-Ramond }
\labelformat{section}{Section #1} 
\labelformat{subsection}{Section #1} 
\labelformat{subsubsection}{Section #1}
\labelformat{subsubsubsection}{Section #1}
\labelformat{equation}{Eq.~(#1)} 
\labelformat{figure}{Fig.~#1} 
\labelformat{subfigure}{Fig.~\thefigure#1} 
\labelformat{table}{Tab.~#1} 
\labelformat{appendix}{Appendix #1}
\title{Horndeski theories confront the Gravity Probe B experiment}
\author{Sajal Mukherjee\footnote{sm13ip029@iiserkol.ac.in}~$^{1}$ and Sumanta Chakraborty\footnote{sumantac.physics@gmail.com}~$^{2}$\\
$^{1}${\small{Department of Physical Sciences, IISER Kolkata, Mohanpur-741246, India}}\\
$^{2}${\small{Department of Theoretical Physics, Indian Association for the Cultivation of Science, Kolkata-700032, India}}}
\begin{document}
  
\maketitle
\begin{abstract}
In this work we have investigated various properties of a spinning gyroscope in the context of Horndeski theories. In particular, we have focused on two specific situations --- (a) when the gyroscope follows a geodesic trajectory and (b) when it is endowed with an acceleration. In both these cases, besides developing the basic formalism, we have also applied the same to understand the motion of a spinning gyroscope in various static and spherically symmetric spacetimes pertaining to Horndeski theories. Starting with the Schwarzschild de-Sitter spacetime as a warm up exercise, we have presented our results for two charged Galileon black holes as well as for a black hole in scalar coupled Einstein-Gauss-Bonnet gravity. In all these cases we have shown that the spinning gyroscope can be used to distinguish black holes from naked singularities. Moreover, using the numerical estimation of the geodetic precession from the Gravity Probe B experiment, we have constrained the gauge/scalar charge of the black holes in 
these Horndeski theories. Implications are also discussed.
\end{abstract}
\section{Introduction}\label{Horn_B_Intro}

General relativity has been very successful in explaining the kinematics as well as the dynamics of our observable universe. The success story of \gr, since its discovery, continues to grow steadily as it passes more and more experimental tests with flying colours. Apart from the earlier predictions made by Einstein, such as bending of light, perihelion precession of Mercury and Gravitational redshift \cite{meaning,carroll2004spacetime,will1993theory,gravitation}, the modern era is also blessed with numerous fruitful tests of \gr. Some notables among them are, the Hulse and Taylor experiment \cite{Hulse:1974eb}, the findings of Gravity Probe B \cite{Everitt:2011hp} and of course the most recent discovery of Gravitational Waves \cite{Abbott:2016blz,Abbott:2016nmj,TheLIGOScientific:2017qsa}. These discoveries have tested \gr\ both in the weak field as well as in the strong field regime. For example, the merger of two black holes is necessarily a strong field phenomena, while, the \GPB\ experiment has been 
carried out in the weak field regime. As the later one is of relevance in the present context, we shall briefly describe the same below. \GPB\ was designed to measure the inertial frame dragging and geodetic precession effect of a spinning gyroscope orbiting Earth due to the Earth's gravitational field. To have a better estimate, four gyroscopes were used and they were all placed in a satellite orbiting the earth at an approximate altitude of $650$ km from the Earth's surface and with an orbital time period of $97.65$ min. The measured values of the geodetic precession and frame dragging by \GPB\ are, $6601.8 \pm 18.3$ mas/yr and $37.2 \pm 7.2$ mas/yr respectively, while \gr\ predicts them to be $6606.1$ mas/yr and $39.2$ mas/yr. This clearly suggests that \gr\ is indeed in close agreement with the experimental evidences \cite{Everitt:2011hp,Silbergleit:2015jrh,Everitt:2015qri,Buchman:2000quk} as far as \GPB\ is concerned.

Beside this enormous triumph, \gr\ has its fair share of limitations as well. It fails both at the very large and at the very small length scales. In particular, \gr\ cannot explain (without invoking some exotic matters like dark energy) the accelerated expansion of the universe \cite{Smoot:1992td,Perlmutter:1998np,Riess:1998cb,Padmanabhan:2002ji} and also it completely breaks down near the singularities, thereby losing its predictive power \cite{Hawking:1976ra}. This suggests that \gr\ behaves as an effective theory and may be replaced by some more fundamental theory at both these scales. This motivates the search for a modified theory of gravitation that can explain (or, better cure) both these shortcomings from a more fundamental level \cite{Clifton:2011jh,Will:2005va,Capozziello:2011et}. The most economic way to achieve the same would be to modify the Einstein-Hilbert action by either incorporating higher curvature terms or by introduction of some additional scalar or tensor fields. Among the 
modifications of 
the \EH action originating from the inclusion of higher curvature terms, a few are of considerable interest. In particular, $f(R)$ theories of gravity has drawn significant interest in the past few years due to its ability to explain the late time cosmic acceleration \cite{Nojiri:2010wj,Sotiriou:2008rp,Nojiri:2003ft,DeFelice:2010aj,Abebe:2013zua,Haghani:2012zq,Starobinsky:1980te,Barrow:1988xh,Maeda:1987xf,Carroll:2003wy,Hu:2007nk,Nojiri:2007as,Capozziello:2006dj} and its close correspondence with scalar-tensor theories of gravity \cite{Capozziello:2010sc,Chakraborty:2016gpg,Chakraborty:2016ydo,Bahamonde:2016wmz,Addazi:2016hip,Guo:2017rul,Sami:2017nhw,Banerjee:2017lxi}. In addition Lovelock theories of gravity \cite{Padmanabhan:2013xyr,Dadhich:2012ma,Dadhich:2008df,Chakraborty:2014rga,Chakraborty:2014joa,Chakraborty:2015wma,Chakraborty:2017zep,Concha:2017nca}, $f(T)$ gravity \cite{Lin:2016nvj,Addazi:2017lat,Chakraborty:2012kj,Capozziello:2001mq,Capozziello:2012zj,Chakraborty:2016lxo}, higher dimensional along 
with higher curvature modifications to gravitational dynamics \cite{Paul:2016itm,Jusufi:2017vew,Chakraborty:2015taq,Chakraborty:2014xla,Chakraborty:2015bja,Chakraborty:2017qve} play crucial roles in explaining various scenarios among the alternative gravity theories. On the other hand, among the scalar coupled gravity theories (also known as the scaler-tensor theories) Horndeski theories are of particular interest. Since for them the field equations are still of second order, no Ostrogradsky ghosts are present \cite{Woodard:2015zca,Horndeski:1974wa,Heisenberg:2014rta,Babichev:2016rlq,Charmousis:2014mia,Yunes:2011we,Rinaldi:2012vy,Pani:2011xm,Iorio:2012cm,Sotiriou:2015lxa,Langlois:2015skt,Crisostomi:2016czh,Cisterna:2016vdx,Cisterna:2015yla,Jimenez:2016isa,Babichev:2015rva,Barausse:2015wia,Valdivia:2017sat,Hou:2017cjy,Bhattacharya:2016naa}. These theories have recently been studied quite extensively in the context of cosmology and black hole physics. Given the importance of Horndeski theories and their 
interesting solution space, it is legitimate to 
ask how the Horndeski theories confront various experimental tests of gravity, thus providing constraints on the parameters of these theories \cite{Clifton:2011jh}. Such an exercise has already been carried out in \cite{Bhattacharya:2016naa}, in the context of perihelion precession of Mercury and bending angle of light (see also \cite{Chakraborty:2012sd}). In this work we will concentrate on the \GPB\ experiment and ask whether it can provide useful constraints on the Horndeski theories. Throughout this work we will use the geometric unit with $c=1$ and $G=1$ unless otherwise stated.

As a first example, let us discuss the charged Galileon black holes, a sub-class of Horndeski theories. Besides non-minimal coupling between scalar and gravity the above model also inherits an additional gauge field which couples to the scalar sector non-minimally. This particular model has been explored earlier in detail \cite{Babichev:2013cya,Cisterna:2014nua,Babichev:2015rva}. The corresponding action for the complete system, including gravity, scalar and gauge field takes the following form \cite{Babichev:2015rva}
\begin{align}
\mathcal{A}=\frac{1}{16\pi}\int d^{4}x~\sqrt{-g}\Bigg[R-\frac{1}{4}F_{\mu \nu}F^{\mu \nu}
&+\beta G^{\mu \nu}\nabla_ {\mu}\Phi \nabla_ {\nu}\Phi-\eta \partial _{\mu}\Phi \partial ^{\mu}\Phi 
\nonumber
\\
&-\frac{\gamma}{2} \left(F_{\mu \sigma}F_{\nu}^{~\sigma}-\frac{1}{4}g_{\mu \nu}F_{\alpha \beta}F^{\alpha \beta}\right)\nabla ^{\mu}\Phi \nabla ^{\nu}\Phi \Bigg]
\end{align}
where the coupling constant $\beta$ is assumed to be non-vanishing. Here the action for the gravity sector is taken to be the \EH term, for the gauge field it is the canonical $-(1/4)F_{\mu \nu}F^{\mu \nu}$ term and finally for the scalar one has the standard kinetic term. However in addition to the above, the theory inhibits two more pieces --- (a) non-minimal coupling of gravity with scalar field through the Einstein tensor $G^{\mu \nu}$ and (b) coupling of the stress tensor of the gauge field to the scalar sector. These two pieces sit in the action with arbitrary dimensionful coefficients $\beta$ and $\gamma$ respectively. Even though the field equations in this simplified setting are complicated, one can use the additional shift-symmetry of the Galileon field $\Phi$, such that $\Phi \rightarrow \Phi +\textrm{constant}$, to derive a conserved Noether current. Imposing spherical symmetry further simplifies the field equations and hence it becomes possible to obtain exact solutions. In the 
case with $\eta=0$, i.e., in absence of any canonical kinetic term for $\Phi$ one obtains the following spherically symmetric solution \cite{Babichev:2015rva}
\begin{align}\label{charge_Gal_BH}
ds^{2}=-\left(1-\frac{2M}{r}+\frac{\gamma(Q^{2}+P^{2})}{4\beta r^{2}}\right)dt^{2}+\left(1-\frac{2M}{r}+\frac{\gamma(Q^{2}+P^{2})}{4\beta r^{2}}\right)^{-1}dr^{2}+r^{2}d\Omega ^{2}
\end{align}
where the charges associated with the gauge field are independent and can be obtained from $F_{tr}=Q/r^{2}$ and $F_{\theta\phi}=P\sin \theta$. Further the scalar field (or, the Galileon field) present in this model takes the following form \cite{Babichev:2015rva}
\begin{align}
\Phi(r)=\Phi_{0}t+\psi(r);\qquad \psi'(r)^{2}=\frac{\frac{2M}{r}-\frac{\gamma(Q^{2}+P^{2})}{4\beta r^{2}}}{\left(1-\frac{2M}{r}+\frac{\gamma(Q^{2}+P^{2})}{4\beta r^{2}}\right)^{2}}\Phi_{0}^{2}
\end{align}
Here the additional constant $\Phi_{0}$ appearing in the solution for the scalar field is related to the coefficient of the non-minimal coupling between Galileon and gravity, as $\Phi_{0}^{2}=1/\beta$. Thus one must have $\beta >0$ to ensure a real solution for $\Phi(r)$. Further the gauge field as well as the scalar field with the positive branch of the above equation for $\psi(r)$ is regular at the event horizon. At this stage one has no conditions on the coupling between the gauge field and the Galileon, thus the sign of $1/r^{2}$ term can have either signs. 

It is also possible to keep $\eta \neq 0$ and hence the relevant static and spherically symmetric solution becomes \cite{Babichev:2015rva}
\begin{align}\label{dS_charge_BH}
ds^{2}=-\left(1-\frac{2M}{r}+\frac{\eta r^{2}}{3\beta}+\frac{\gamma(Q^{2}+P^{2})}{4\beta r^{2}}\right)dt^{2}+\left(1-\frac{2M}{r}+\frac{\eta r^{2}}{3\beta}+\frac{\gamma(Q^{2}+P^{2})}{4\beta r^{2}}\right)^{-1}dr^{2}+r^{2}d\Omega ^{2}
\end{align}
where the ratio $\eta/3\beta$ acts as the negative of the effective cosmological constant. In order to be consistent with the accelerated expansion of the universe at the large scale we consider the de-Sitter branch of the above solution, which requires $\eta <0$. In this case as well the Galileon field and its derivative are regular at the event horizon. However in this case the electric and magnetic charges are not independent and one must have $\gamma >\beta >0$ to ensure consistent description of the spacetime. 

The final static and spherically symmetric spacetime we will consider is a solution to the scalar coupled Einstein-Gauss-Bonnet gravity and corresponds to a spherically symmetric black hole solution with scalar hair. We will refer to this solution as the Sotiriou-Zhau solution \cite{Sotiriou:2013qea,Sotiriou:2014pfa} (also see \cite{Campbell:1990ai,Campbell:1990fu,Duncan:1992vz,Kanti:1995vq,Antoniou:2017acq}). The above solution is derived assuming a linear coupling between the scalar field $\Phi$ and the Gauss-Bonnet invariant $L_{\rm GB}$, such that the action becomes \cite{Kobayashi:2011nu,Sotiriou:2014pfa}
\begin{equation}
\mathcal{A}=\frac{1}{8\pi}\int d^{4}x\sqrt{-g}~\left(\frac{R}{2}-\frac{1}{2}\partial _{\mu}\Phi \partial ^{\mu}\Phi +\alpha \Phi L_{\rm GB}\right)
\end{equation}
The Gauss-Bonnet invariant appearing in the above action can be written in terms of various curvature quantities and has the following expression: $L_{\rm GB}=R^2-4R^{\alpha \beta}R_{\alpha \beta}+R^{\alpha \beta \mu \nu}R_{\alpha \beta \mu \nu}$, with $R$, $R^{\alpha \beta}$ and $R^{\alpha \beta \mu \nu}$ having the usual meaning of Ricci scalar, Ricci tensor and Riemann curvature tensor respectively. The metric associated with the hairy black hole in Einstein-Gauss-Bonnet gravity correspond to
\begin{equation}
ds^2 =-f(r)dt^2+h(r)dr^2+r^2 d\Omega^2~,
\label{Metric_SZ}
\end{equation}
where the metric elements $h(r)$, $f(r)$ as well as the scalar field profile $\Phi(r)$ reads \cite{Sotiriou:2013qea,Sotiriou:2014pfa}:
\begin{eqnarray}
f(r)& \approx & 1-\dfrac{2M}{r}+\dfrac{MP^2}{6 r^3}+\mathcal{O}(r^{-4}); \qquad h(r) \approx 1+\dfrac{2M}{r}+\dfrac{8 M^2-P^2}{2 r^2}+\mathcal{O}(r^{-3})~,\nonumber \\
\Phi(r) & \approx & \dfrac{P}{r}+\dfrac{M P^2}{r^2}+\mathcal{O}(r^{-3})~.
\label{eq:SZ_Appx}
\end{eqnarray}
Here $P$ is the scalar charge (or, scalar hair) associated with the above black hole. Further note that the field equation for the scalar field correspond to $\square \Phi +L_{\rm GB}=0$, which can be trivially integrated, since in four dimensions the Gauss-Bonnet invariant is a total derivative term. As a consequence, one can demonstrate that the above solution can \emph{not} represent the exterior geometry of a compact object, rather can only depict a black hole spacetime. In the present context, we shall explicitly use \ref{eq:SZ_Appx} and find out the features associated with the motion of a gyroscope in this hairy black hole spacetime. In addition to the above two solutions, we will also consider the Schwarzschild de-Sitter solution to set the stage for the charged Galileon black hole and the Sotiriou-Zhou solutions. 

The paper is organized as follows: In \ref{Precesion_Geodesic}, we derive the geodetic precession in a general static and spherically symmetric spacetime, which subsequently have been extended for a gyroscope on a non-geodetic trajectory and have computed its precession frequency in \ref{Precession_NonGeodesic}. The techniques developed in the earlier sections have been applied in \ref{Horndeski_Application} to study the motion of a gyroscope in both geodesic and accelerated trajectories for spacetimes originating from Horndeski theories. Moreover we have explicitly pointed out the features associated with these Horndeski theories, but are absent in \gr. We have also discussed the viability of these theories and future directions of exploration \'{a} la the \GPB\ experiment.
\section{Geodetic Precession in a general static, spherically symmetric spacetime: Formalism}\label{Precesion_Geodesic}

In this section we will discuss in detail the geodetic precession of a spinning gyroscope in a general static and spherically symmetric spacetime. The presence of an external static and spherically symmetric gravitational field will lead to precession of the spinning axis of the gyroscope, which we will compute in our general framework. This will enable us to evaluate the precession angle of the gyroscope for spherically symmetric spacetimes in gravity theories other than \gr. Thereby one can easily read off the effect of these alternative gravity theories on the precession frequency. This in turn possibly can be used to provide stringent bounds on the parameters appearing in these alternative gravity models using the results from Gravity Probe B.

Given the above motivation we will now concentrate on the derivation of the precession angle for a gyroscope moving in a circular geodesic in the static and spherically symmetric spacetime. Later on we will also discuss the situation for accelerated gyroscopes as well. Since the spacetime is static and spherically symmetric there exists two obvious Killing vectors, namely $t^{a}=\left(\partial/\partial t\right)^{a}$, defining a Killing time $t$ and $\phi^{a}=\left(\partial/\partial \phi \right)^{a}$, defining an appropriate angular coordinate $\phi$. Thus in these Killing coordinates the general static and spherically symmetric spacetime takes the following form,
\begin{equation}
ds^2=-e^{\nu(r)} dt^2+e^{\lambda(r)}dr^2+r^2 d\Omega^2~.
\label{eq:Metric}
\end{equation}
Here $\nu(r)$ and $\lambda(r)$ are arbitrary function of the radial coordinate only, to be determined from the field equations of gravity. Any modification of the gravitational action over and above general relativity will results into modified field equations and hence will modify the functions $\nu(r)$ and $\lambda (r)$ as well. This in turn will lead to corrections in the geodetic precession in comparison to general relativity, which we will explore in this work.

As the above solution depicts a spherically symmetric spacetime, we can comfortably choose to work on the equatorial slice, i.e., with $\theta = \pi/2$ without losing any generality. The four-velocity of a gyroscope on a circular trajectory in the equatorial plane must satisfy the following conditions: $U^{r}=dr/d\tau=0=U^{\theta}=d\theta/d\tau$, where $\tau$ is the proper time along the circular trajectory. Thus both the radial and $\theta$ component of the four-velocity must vanish. Further if $\Omega _{\rm g}$ is the angular velocity of the observer on the circular geodesic then, $\Omega _{\rm g}=d\phi/dt=U^{\phi}/U^{t}$. This enables us to write down the four-velocity of the gyroscope moving in a circular geodesic as
\begin{equation}\label{four_velocity_geo}
U^a =N\left(1,0,0,\Omega_{\rm g}\right)~,
\end{equation}
where, $N$ is the overall normalization factor. To determine $\Omega _{\rm g}$, we need to know the energy and angular momentum associated with the gyroscope, for which we assume that the radius of the circular orbit is $r_{c}$. Given this information, the energy $E_{c}$ and angular momentum $L_{c}$ associated with the gyroscope become \cite{Chakraborty:2015vla}
\begin{equation}
E_{c}^{2}=\dfrac{2e^{\nu_{c}}}{2-r_{c}\nu'_{c}} \qquad L_{c}^{2}=\dfrac{r_{c}^{3}\nu'_{c}}{2-r_{c}\nu'_{c}}~.
\end{equation}
In the above expressions for energy and angular momentum, `prime' denotes derivative with respect to the radial coordinate and $\nu_{c}\equiv \nu(r_{c})$ along with $\nu'_{c}\equiv d\nu/dr$ evaluated at $r=r_{c}$. Thus the angular velocity associated with the gyroscope in the circular geodesic takes the following form,
\begin{equation}\label{Ang_Vel_Geo}
\Omega^2_g = (U^{\phi}/U^t)^2 = \dfrac{e^{2\nu_{c}}}{r_{c}^4}\left(\dfrac{L_{c}}{E_{c}}\right)^2=\dfrac{e^{\nu_{c}}\nu'_{c}}{2r_{c}}~.
\end{equation}
Having derived the angular velocity it is easy to determine the overall normalization factor by solving $U^{a}U_{a}=-1$, which in view of \ref{four_velocity_geo} takes the following form: $-e^{\nu_{c}}N^{2}+r_{c}^{2}N^{2}\Omega _{\rm g}^{2}=-1$. Thus using the expression for $\Omega _{g}^{2}$ from \ref{Ang_Vel_Geo} the normalization factor becomes, $N^{2}=2\exp(-\nu _{c})/(2-r_{c}\nu'_{c})$. 

The expression for the normalization factor diverges as $r_{c}\rightarrow r_{\rm ph}$, where $r_{\rm ph}$ is the photon circular orbit. This is because the photon circular orbit $r_{\rm ph}$ satisfies the following differential equation $2=r_{\rm ph}\nu'_{\rm ph}$. It is expected since, there can be no timelike observer moving on a circular orbit with radius $r_{\rm ph}$. This tells us that the spacetime region beyond $r=r_{\rm ph}$ is not accessible to observers moving in a circular geodesic.  We will discuss the corresponding situation for non-geodesic gyroscopes in the later sections. The above completes our discussion as far as the motion of the gyroscope in a circular geodesic is concerned, we will now concentrate on the evolution of the spin of the gyroscope as it moves along the circular geodesic.

The spin of the gyroscope will be described by the spin four-vector $S^{a}$, which is orthogonal to the velocity four-vector $U^{a}$, such that $S^{a}U_{a}=0$. Further spin four-vector will change as the gyroscope moves along the circular geodesic and the rate of change of the spin four-vector corresponds to $dS^{a}/d\tau$. This can be achieved by using the fact that the spin four-vector is parallel transported along the circular geodesic, such that
\begin{equation}
\dfrac{DS^a}{d\tau}=\frac{dS^{a}}{d\tau}+\Gamma ^{a}_{bc}U^{b}S^{c}=0~.
\label{eq_Spin}
\end{equation}
This equation will be used to determine the spin four-vector after the gyroscope has made one complete rotation, using which one can evaluate the spin precession. Before that, we can use the orthogonality condition: $U^{a}S_{a}=0$, along with \ref{four_velocity_geo} to arrive at $-\exp(\nu_{c})NS^{t}+r_{c}^{2}N\Omega _{\rm g}S^{\phi}=0$. From which it is straightforward to determine the temporal component of the spin four-vector in terms of the angular part as,
\begin{equation}\label{spin_rel}
S^{t}=r_{c}^{2}\Omega _{\rm g}e^{-\nu_{c}}S^{\phi}~.
\end{equation}
Having derived the above relation one can now use the evolution equation for the spin four-vector using the affine connections for the spherically symmetric metric and \ref{eq_Spin} to arrive at the following differential equations \cite{padmanabhan2010gravitation},
\begin{equation}
\frac{dS^{t}}{d\tau}=-Nr_{c}e^{-\nu_{c}}\Omega _{\rm g}^{2}S^{r};\qquad 
\frac{dS^{r}}{d\tau}=e^{-\lambda _{c}-\nu_{c}}\frac{r_{c}\Omega _{\rm g}}{N}S^{\phi};\qquad
\frac{dS^{\phi}}{d\tau}+\frac{N\Omega _{\rm g}}{r_{c}}S^{r}=0;\qquad
\frac{dS^{\theta}}{d\tau}=0~.
\end{equation}
One can use the relation between $S^{t}$ and $S^{\phi}$ from \ref{spin_rel} to eliminate $S^{t}$ from the above equations. Moreover differentiating the above expressions again with respect to the proper time $\tau$ and converting $\tau \rightarrow t$ using the relation: $dt=Nd\tau$ we finally obtain the following evolution equations,
\begin{equation}\label{Evolution_Equations}
\dfrac{d^2S^r}{dt^2}+e^{-(\nu_{c}+\lambda_{c})} \left(\dfrac{\Omega_g}{N}\right)^2 S^r=0; \qquad 
\dfrac{dS^{\phi}}{dt}=-\dfrac{\Omega_g}{r_{c}} S^{r}~.
\end{equation}
Thus the $S^{r}$ component of the spin four-vector satisfies the differential equation of a simple harmonic oscillator, which can be solved to yield: $S^{r}(t)=S^{r}(0)\cos \omega_{\rm g}t$. Here we have imposed the initial conditions such that the spin three-vector was initially directed along the radial direction, i.e., $S^{\theta}(0)=0=S^{\phi}(0)$. Given this solution for the radial component one can immediately solve for $S^{\phi}$, yielding: $S^{\phi}(t)=-(\Omega _{\rm g}S^{r}(0)/r_{c}\omega_{\rm g})\sin \omega _{g}t$. In both these solutions for $S^{r}(t)$ and $S^{\phi}(t)$ we have introduced a new frequency of oscillation pertaining to the spin four-vector defined as,
\begin{equation}\label{spin_rot}
\omega _{\rm g}\equiv \Omega _{\rm g}\left[e^{-\lambda _{c}}\left(\frac{2-r_{c}\nu'_{c}}{2}\right)\right]^{1/2}~.
\end{equation}
Note that for Schwarzschild solution, $e^{-\lambda}=e^{\nu}=1-(2M/r)$, the term inside square root becomes $1-(3M/r)$ which coincides exactly with the earlier literatures \cite{gravitation}. However we would like to stress that the above expression is completely general, given any static and spherically symmetric spacetime one can directly employ the results derived above. Further note that, the sign of $S^{\phi}$ is negative, which tells us that both the $S^{r}$ and $S^{\phi}$ components of the spin rotate relative to the initial radial direction with an angular velocity $\omega _{\rm g}$. However as evident from \ref{spin_rot} the angular velocity of rotation of the spin three vector is different from the angular velocity of rotation of the gyroscope along the circular trajectory, resulting in spin precession. That is, when the gyroscope completes one rotation along the circular geodesic, the spin three-vector has not yet completed a complete rotation as $\omega_{\rm g}<\Omega _{\rm g}$. This results into 
a precession of the spin three-vector, which is called \emph{geodetic precession} and for one complete revolution of the gyroscope along the circular orbit, it is given by,
\begin{equation}\label{Geod_prec}
\mathcal{G}_{\rm g}=2\pi\left(1-\dfrac{\omega_g}{\Omega_g}\right)
=2\pi \left(1-\sqrt{\dfrac{2-r_{c}\nu'_{c}}{2 e^{\lambda_{c}}}}\right)~.
\end{equation}
The above expression for geodetic precession produces appropriate Schwarzschild limit as one can easily verify. Furthermore, we would like to point out that the geodetic precession becomes $2\pi$ on the circular photon orbit $r_{\rm ph}$. This merely points out that there can be no timelike circular geodesic on $r_{\rm ph}$. For Schwarzschild solution it turns out that this geodetic precession vanishes only for $r\rightarrow \infty$ and does not vanish for any finite $r$. Thus it would be of interest to explore if the geodetic precession can vanish at any finite radial distance if one considers alternative gravity theories. This will provide a very nice discriminating feature of these alternative theories, setting them apart from \gr. Besides a numerical computation of the geodetic precession for near-earth artificial satellites makes it possible to constrain parameters in these alternative theories using \GPB. Taking the gyroscope to be an artificial satellite spaced at an attitude of $650$ km and having a 
orbital time period of $97.65$ min, one arrives at an geodetic precession which is $6606.1$ mas per year. On the other hand, \GPB\ 
has measured the geodetic precession of such a system to be $6601.8$ mas per year with an error of $\pm 18.3$ mas. Thus any deviation of the geodetic precession from general relativity should fall within the above error bound. With this information about geodetic precession in the backdrop, we can compute the same in various alternative theories and see what constraints these alternative theories should satisfy so that the geodetic precession falls with the error bound prescribed by \GPB. This is exactly what we will try to provide in the later parts of this work. 
\section{Precession for non-geodesic observers: The Frenet-Serret formalism}\label{Precession_NonGeodesic}

For a gyroscope moving in a non-geodesic trajectory, there exist a formalism known as the \FS formalism in order to compute the precession of the gyroscope \cite{Iyer:1993qa} (however also see, \cite{Deriglazov:2015zta,Deriglazov:2017jub}). This formalism requires spacetime to inherit certain symmetries and hence calls for the existence of Killing vectors. One assumes that the trajectory of the gyroscope corresponds to a quasi-Killing orbit, i.e., the four-velocity $U^{a}$ is a linear combination of the Killing vectors associated with the spacetime. The worldlines of the observer can be determined by three scalars --- (a) $\kappa$, known as curvature along the curve and (b) $\tau _{1}$ and $\tau _{2}$, representing the two torsion parameters along the curve. These three scalars can be derived in terms of the quasi-Killing vector $U^{a}$ and its various derivatives.

Given this set up one can use the following equation depicting the evolution of the spin vector $S^{a}$ for a Fermi dragged gyroscope,
\begin{equation}
\dfrac{DS^{a}}{d\tau}=\left(S^{b}a_{b}\right)U^a~.
\label{eq:fermi_dragged}
\end{equation}
where we have used the fact that the spin vector is orthogonal to the velocity four-vector, i.e., $S^{a}U_{a}=0$. In the above expression one notes that $a^{i}=U^{a}\nabla _{a}U^{i}$, is the acceleration of the spinning gyroscope. In the case of geodesic motion the acceleration identically vanishes and hence \ref{eq_Spin} follows. 

We would like to emphasize that the above equation is the evolution equation for the spin of a gyroscope along the trajectory of the observer and is a particular form of the Mathisson-Papapetrau equations \cite{Mathisson:1937zz, Papapetrou:1951pa,Corinaldesi:1951pb}. These equations describe the motion of a pole-dipole particle in a gravitational field and can be solved only under spin supplementary conditions. In a more technical way, $S^aU_a=0$ is referred to as a particular constraint namely the Mathisson-Pirani spin supplementary condition and is extremely important in the context of spinning particles. There exist two different ways to compute the spin precession frequency of the gyroscope starting from \ref{eq:fermi_dragged}, one of them has been discussed in detail in \cite{straumann2009general} and used explicitly in \cite{Chakraborty:2016mhx}, while the other method is described in \cite{Iyer:1993qa,Nayak:1998er}. We would closely follow the second approach in this work to determine the spin 
frequency of a gyroscope moving along a non-geodesic trajectory. 

The trajectory of the gyroscope is defined as the \FS frame and is determined by the following set of orthonormal tetrads denoted by $\{e^{a}_{(0)},e^{a}_{(\alpha)}\}$, where $\alpha$ denotes the spacelike components and $e^{a}_{(0)}=U^{a}$. On the other hand, the spin of the gyroscope is being Fermi-Walker transported along the trajectory and hence can be defined using a separate orthogonal tetrad $\{e^a_{(0)},f^a_{(\alpha)}\}$. Given the two different tetrads, the gyroscopic precession can be understood as the relative angular velocity between these two frames and hence takes the following form \cite{Iyer:1993qa}
\begin{equation}
\Omega^a_{\rm FS}= \tau_1 e^a_{(3)}+\tau_2 e^a_{(1)}~. 
\end{equation}  
Where, $\tau_1$ and $\tau_2$ are the two torsion parameters. For a gyroscope moving in a circular orbit of radius $r_{\rm c}$ on the equatorial plane of a static and spherically symmetric spacetime with constant angular velocity $\Omega _{\rm nongd}$, it follows that $\tau _{2}=0$ and hence only $\tau _{1}$ survives. Since the vector $e^{a}_{(3)}$ is orthonormal it follows that the magnitude of precession frequency to be $\tau _{1}$. For such a gyroscope it follows that the torsion parameter takes the following value \cite{Iyer:1993qa},
\begin{equation}\label{torsion}
\tau_1 = g^{rr}_{c}\Omega _{\rm nongd} \left(1-r_{c}a_{c}\right)e^{(\lambda _{c}-\nu_{c})/2}~,
\end{equation}
where any quantity with subscript `c' implies that it has been evaluated on the circular trajectory located at $r=r_{c}$. The radial acceleration of the particle is being denoted by $a_{c}$ (i.e., $a(r_{c})$) and has the following expression,
\begin{equation}
a_{c}=\dfrac{\nu'_{c}e^{\nu_{c}}-2 r_{c} \Omega_{\rm nongd}^2}{2 (e^{\nu_{c}}-r_{c}^2\Omega_{\rm nongd}^2)}~.
\label{eq:acceleration}
\end{equation}
Using the radial acceleration from \ref{eq:acceleration} in the expression for the first torsion parameter $\tau _{1}$ in \ref{torsion}, we arrive at,
\begin{equation}\label{FD_Rot}
\Omega _{\rm FS}=\tau_{1}=\dfrac{\Omega_{\rm nongd}}{2}e^{\left(\nu_{c}-\lambda_{c}\right)/2} \left(\dfrac{2-r_{c}\nu'_{c}}{e^{\nu_{c}}-r_{c}^2\Omega_{\rm nongd}^2}\right)~.
\end{equation} 
One can immediately check that the above expression matches exactly with the results obtained in \cite{Chakraborty:2016mhx}. Note that $\Omega _{\rm FS}$ is the precession frequency of the gyroscope with respect to the local inertial frame, i.e., it is defined with respect to the proper time along the trajectory of the observer. However we would like to convert the same to the Killing time coordinate $t$ and thereby introducing a redshift factor. Thus the precession frequency $\omega _{\rm nongd}$ for non-geodesic observer becomes,
\begin{equation}
\omega _{\rm nongd}=\Omega_{\rm FS}\left(\dfrac{dt}{d\tau}\right)_{c}^{-1} = \dfrac{\Omega _{\rm nongd}}{2}e^{\left(\nu_{c}-\lambda_{c} \right)/2} \left(\dfrac{2-r_{c}\nu_{c}'}{\sqrt{e^{\nu_{c}}-r_{c}^2\Omega _{\rm nongd}^2}}\right)~,
\label{Eq:Fermi_dragged}
\end{equation}
where the factor $dt/d\tau$ corresponds to $(e^{\nu}-r^{2}\Omega _{\rm nongd}^{2})^{1/2}$ obtained by the normalization of $U^{a}$. Thus the precession frequency of the gyroscope moving on a circular, but non-geodesic trajectory with constant angular velocity $\Omega _{\rm nongd}$ becomes,
\begin{equation}\label{nongeod_prec}
\mathcal{G}_{\rm nongd}=2\pi \left(1-\dfrac{\omega_{\rm nongd}}{\Omega_{\rm nongd}}\right)=2\pi \left(1-e^{-\lambda_{c}/2}\dfrac{1-\dfrac{r_{c}\nu'_{c}}{2}}{\sqrt{1-r_{c}^2 \Omega _{\rm nongd}^2 e^{-\nu_{c}}}}\right)~.
\end{equation}
Note that as $\Omega_{\rm nongd}$ is being replaced by $\Omega _{\rm g}$, the angular velocity for the geodesic observers, one immediately arrives at \ref{Geod_prec} representing precession of geodesic observers. 

However in this particular situation, unlike the case for geodesic observers, the angular velocity of the non-geodesic observer, $\Omega_{\rm nongd}$ is arbitrary, since there exists no general prescription to write it down. For this purpose, following \cite{Chakraborty:2016mhx} we will express $\Omega$ as a sum of upper and lower bound of photon's angular velocity, which correspond to: $\Omega_{\rm ph}^{\pm}=\pm(e^{\nu_{c}/2}/r_{c})$. Here the plus sign corresponds to rotation in anti-clockwise direction, while the negative sign signifies motion in a clockwise direction. From \ref{FD_Rot} it follows that $\Omega _{\rm FS}$ diverges at $\Omega _{\rm nongd}=\Omega _{\rm ph,nongd}^{\pm}$ and for timelike observer it is essential that $\Omega _{\rm ph, nongd}^{-}<\Omega _{\rm nongd} <\Omega _{\rm ph,nongd}^{+}$. Thus one may define a parameter $\epsilon$, running from $0$ to $1$ and hence set the angular velocity of the observer at a circular orbit of radius $r_{c}$, such that
\begin{equation}\label{nongeod_freq}
\Omega _{\rm nongd}=\epsilon \Omega_{\rm ph,nongd}^{+} +(1-\epsilon) \Omega_{\rm ph,nongd}^{-} = (2\epsilon-1)\sqrt{\dfrac{e^{\nu_{c}}}{r_{c}^{2}}}~.
\end{equation}
Hence one immediately observes $e^{\nu_{c}}-r_{c}^{2}\Omega _{\rm nongd}^{2} =e^{\nu_{c}}\{1-(2\epsilon-1)^{2}\}=4\epsilon (1-\epsilon)e^{\nu_{c}}$. We will use this result extensively later on. 

In this work we will discuss both the geodesic and non-geodesic precession of gyroscopes in alternative theories, with possible interesting phenomenon originating from non-geodesic motion, which in principle can tell us about the underlying structure of the spacetime. From the observational point of view it is difficult to obtain any constraint using the expression for precession along a non-geodesic trajectory, since the Gravity Probe B experiment has been carried out using geodesic trajectory. Hence only the results presented in \ref{Precesion_Geodesic} will be relevant for imposing constraints on the various alternative models of interest, which we will perform in the later sections. Despite the difficulties in measuring the non-geodesic part of the precession, there could be few situations where it becomes important. One such scenario may correspond to determination of magnetic moment of muon, i.e., through muon $g-2$ measurements. However the corresponding effect of non-geodesic 
precession in measurement of muon magnetic moment seems to be quite small \cite{Nikolic:2018dap,Morishima:2018bqz,Laszlo:2018llb}. On the other hand, measurements of electric dipole moment using the frozen spin method do inhibit non-trivial \gr\ corrections with non-geodesic spin precession playing a key role \cite{Laszlo:2018llb}. We leave these non-trivial effects due to non-geodesic motion of spinning particles for future. 
\section{Spin precession in Horndeski theories and \GPB}\label{Horndeski_Application}

In this section we will describe the motion of a spinning gyroscope either on a geodesic or non-geodesic circular trajectory in spherically symmetric spacetimes. These spherically symmetric spacetimes are taken to be solutions of various alternative theories having their origin in one way or another into the Horndeski theories of gravity. The alternative gravity models along with the associated static and spherically symmetric spacetimes have already been discussed in \ref{Horn_B_Intro} and hence we will mainly concentrate on the precession frequencies, namely $\mathcal{G}_{\rm g}$ and $\mathcal{G}_{\rm nongd}$ respectively, in these spacetimes. As a warm up, we will first discuss the effect of cosmological constant on the precession frequencies before taking up the effect of alternative theories on them.  
\subsection{Warm up: Einstein gravity with cosmological constant}

\begin{figure}
\subfloat[The above figure illustrates the variation of the geodetic spin frequency with the radial distance for various choices of the cosmological constant $\Lambda$. It is clear that the spin frequency ($\omega_{\rm g}$) vanishes at two points, one is at the photon circular orbit $r_{\rm ph}=3M$ and another is at $r_{0}=(3M/\Lambda)^{1/3}$. The later one is well inside the cosmological horizon $r_{\rm ch}$ and hence is in principle detectable. \label{Fig_02a}]{\includegraphics[scale=0.5]{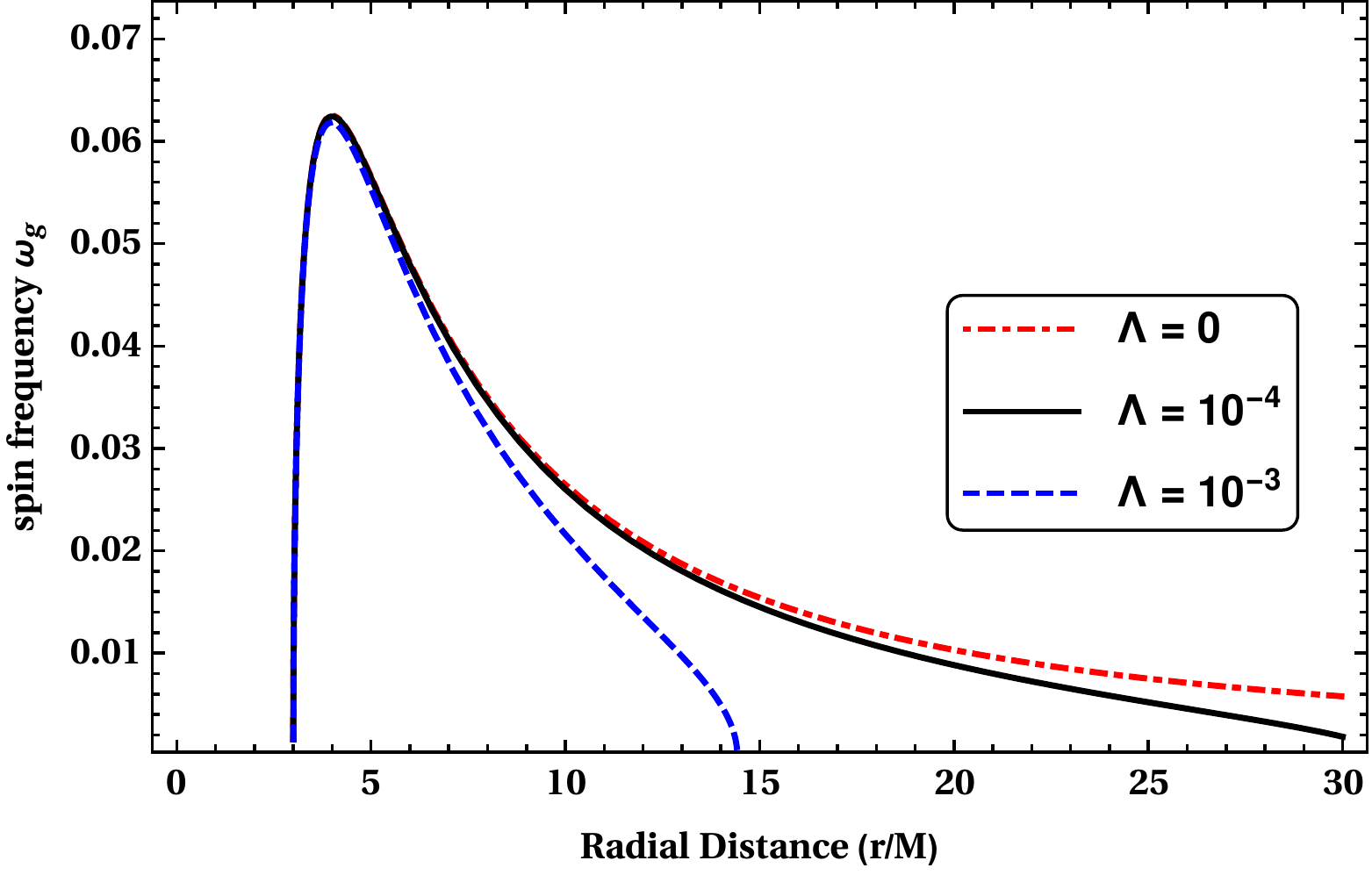}}
\hfill
\subfloat[Spin precession ($\omega_{\rm nongd}$) for a gyroscope moving in a non geodesic circular trajectory is shown for different $\epsilon$ values, with $\Lambda$ being fixed at $10^{-4}M^{-2}$. The frequency $\Omega_{\rm nongd}$ only vanishes at photon circular orbit ($r_{\rm ph}=3M$). Further, for $\epsilon>0.5$, $\omega _{\rm nongd}$ is negative and for $\epsilon<0.5$, $\omega _{\rm nongd}$ is positive.\label{Fig_02b}]{\includegraphics[scale=0.5]{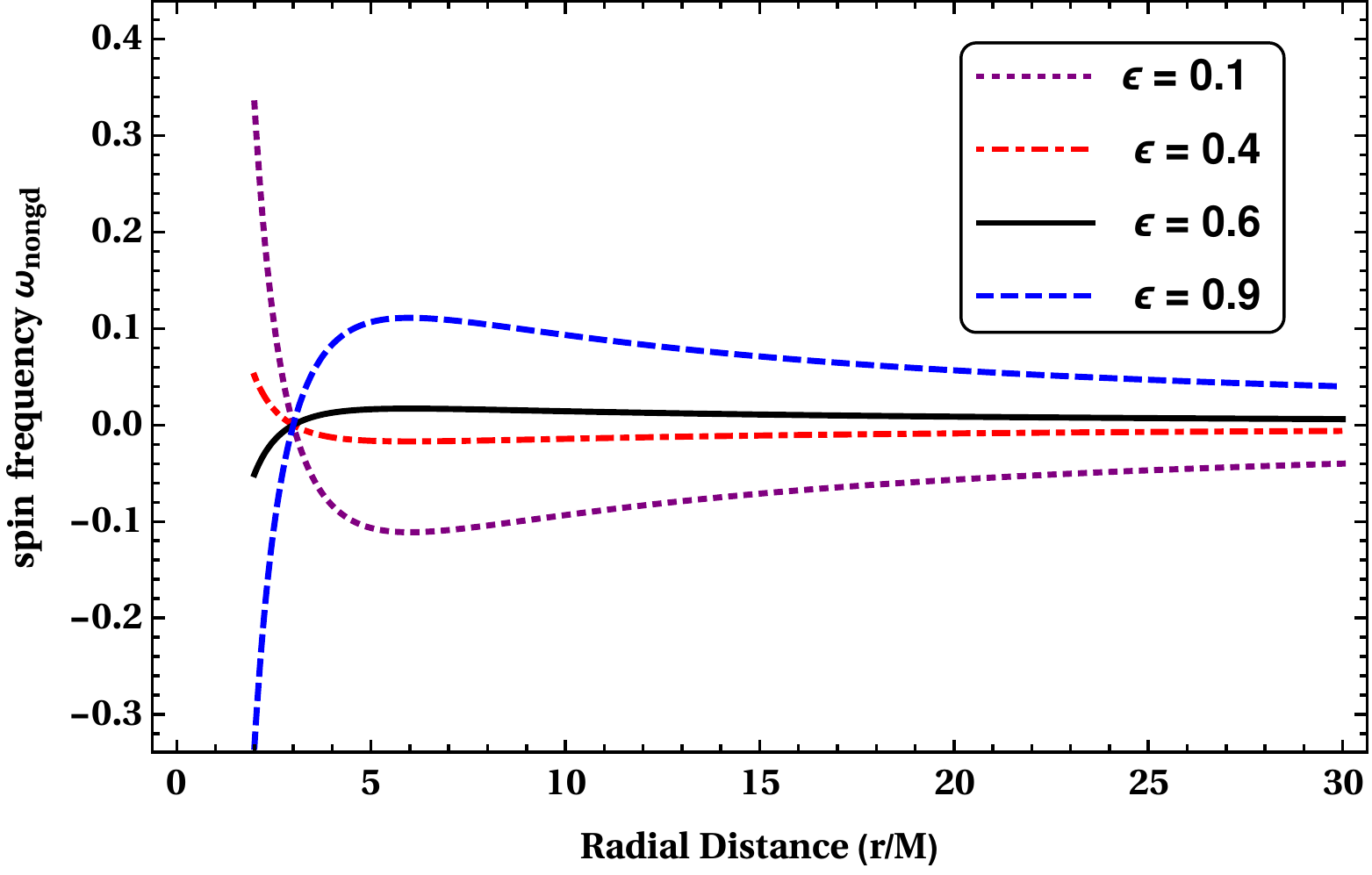}}
\caption{Spin frequency $\omega _{\rm g}$ and $\omega _{\rm nongd}$ for geodesic and non-geodesic gyroscopes respectively have been illustrated in the context of Schwarzschild de-Sitter spacetime.}
\label{Fig_02}
\end{figure}

Before delving into the computation of precession frequencies for alternative theories it is instructive to discuss a more basic scenario as a warm up example. Since the results of geodesic (as well as non-geodesic) precession are well known for Schwarzschild spacetime we consider here the effect of cosmological constant, i.e., precession frequency of a gyroscope in Schwarzschild de-Sitter spacetime. The line element for Schwarzschild-de Sitter spacetime takes the form of \ref{eq:Metric} with $e^{\nu}=1-(2M/r)-(\Lambda/3)r^{2}=e^{-\lambda}$. There exist two horizons in this spacetime, the inner one is the event horizon, while the outer one corresponds to the cosmological horizon. The locations of these horizons can be determined by solving the cubic equation $re^{-\lambda}=0=r-2M-(\Lambda/3)r^{3}$ and hence we obtain 
\begin{align}
r_{\rm eh}&=\frac{2}{\sqrt{\Lambda}}\cos\left[\frac{1}{3}\cos^{-1}\left({3M \sqrt{\Lambda}}\right)+\frac{\pi}{3}\right]~,
\nonumber
\\
r_{\rm ch}&=\frac{2}{\sqrt{\Lambda}}\cos\left[\frac{1}{3}\cos^{-1}\left({3M \sqrt{\Lambda}}\right)-\frac{\pi}{3}\right]~.
\end{align}
In principle, the above cubic equation could have three real roots. However with positive $M$ and $\Lambda$, along with the choice $3M\sqrt{\Lambda}<1$ one arrives at $r_{\rm eh}<r_{\rm ch}$, while the other root becomes negative and hence can be discarded. Here, $r_{\rm eh}$ corresponds to the event horizon and $r_{\rm ch}$ stands for the cosmological horizon. Since $e^{\nu}=e^{-\lambda}$ it is straightforward to compute the angular velocity of the gyroscope moving in a circular geodesic of radius $r_{c}$ using \ref{Ang_Vel_Geo} leading to $\Omega _{\rm g}=\{(M/r_{c}^{3})-(\Lambda/3)\}^{1/2}$. Thus subsequently using \ref{spin_rot} one obtains the spin frequency of the gyroscope to yield,
\begin{equation}
\omega_{g}=\Omega_{\rm g}\sqrt{1-\dfrac{3M}{r_{c}}}=\sqrt{\dfrac{M}{r_{c}^3}-\dfrac{\Lambda}{3}}\sqrt{1-\dfrac{3M}{r_{c}}}
\end{equation}
\begin{figure*}
\begin{center}

\includegraphics[scale=0.7]{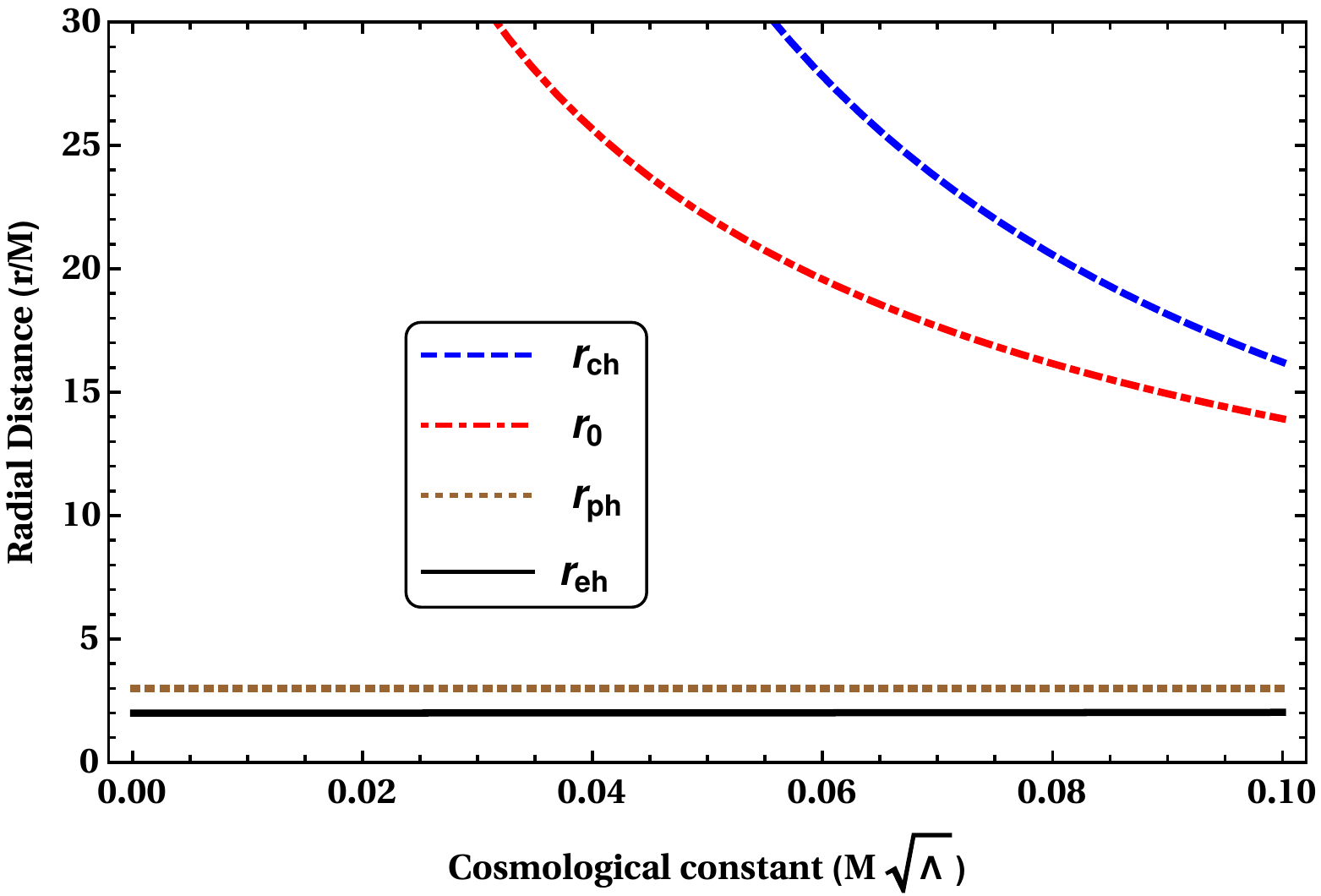}
\caption{The above figure depicts the horizon structure in Schwarzschild de-Sitter spacetime. The event horizon $r_{\rm eh}$ (thick black line), the photon radius $r_{\rm ph}$ (densely dotted brown line), the cosmological horizon $r_{\rm ch}$ (thick dotted blue line) and finally the radius $r_{0}$ (dot dashed red line) have been presented. As the figure clearly demonstrates, the radius $r_{0}$ where the geodetic precession frequency vanishes lies within the photon radius and the cosmological horizon. Thus it is accessible to any stationary observer and can act as a discriminator between Schwarzschild and Schwarzschild de-Sitter spacetime.}
\label{Fig_01}
\end{center}
\end{figure*}
The above expression for spin frequency has some interesting features, namely the ratio $\omega_{g}/\Omega _{g}$ is independent of the cosmological constant $\Lambda$ and coincides with the corresponding expression for Schwarzschild spacetime. As a consequence the geodetic precession frequency becomes, $\mathcal{G}_{g}=2\pi[1-\{1-(3M/r_{c})\}^{1/2}]$. Further, the precession frequency besides vanishing at the usual photon circular orbit ($r_{\rm ph}=3M$) also vanishes at $r_{0}=(3M/\Lambda)^{1/3}$. Thus in order to obtain a non-trivial $\omega _{g}$ it is necessary that the radius of the circular geodesic must satisfy the following criteria: $r_{c}>\textrm{max}(r_{\rm ph},r_{0})$. Moreover, in order to have some observable consequences it is necessary for the radius $r_{0}$ to be located outside the photon circular orbit but within the cosmological horizon. The condition $r_{0}>3M$, requires $\Lambda^{-1}>9M^{2}$, i.e., $\Lambda$ has to be tiny, which for solar mass black holes are trivially satisfied. With 
the above condition imposed on $\Lambda$ one can also ensure that $r_{0}<r_{\rm ch}$ (see \ref{Fig_01}). Thus if one observes that the spin frequency of a gyroscope moving on a circular geodesic is vanishing at some finite radius outside $r_{\rm ph}$, then one may conclude that the spacetime inherits a cosmological constant! Interestingly, given this radius one can also determine the numerical value of the cosmological constant if the black hole mass is known (this has been explicitly demonstrated in \ref{Fig_02a}). 

Since the spin frequency vanishes at $r_{c}=r_{\rm ph}$ and at $r_{c}=r_{0}$, it follows that it must attain a maximum value within this range, which can be obtained by setting $\partial _{r_{c}}\omega _{g}=0$ and then verifying $\partial _{r_{c}}^{2}\omega _{g}>0$. This essentially corresponds to the real root of the cubic equation, $\Lambda r_{c}^{3}+3r_{c}-12M=0$ and has also been clearly illustrated in \ref{Fig_02a}. Note that there is very little effect of the numerical value of the cosmological constant on the location of the maxima, which is approximately situated at, $r_{c}\simeq 4M$. 

We would like to emphasize that this feature is very much unique and appears in the presence of cosmological constant alone. In normal Schwarzschild black hole there will be a maxima, but it will finally go to zero at infinity. While in Schwarzschild de-Sitter spacetime it follows that the spin  frequency has to vanish at some finite radius before the cosmological horizon. This serves as a distinct feature of Schwarzschild de-Sitter spacetime.
Before finishing this section, let us briefly comment on the precession frequency for non-geodesic observer, using the Frenet-Serret formalism developed in \ref{Precession_NonGeodesic}. Given the metric elements for the Schwarzschild de-Sitter spacetime, one can immediately compute the spin frequency $\omega _{\rm nongd}$ using \ref{nongeod_freq} and \ref{Eq:Fermi_dragged} respectively,
\begin{equation}\label{Eq_dS_ng}
\omega_{\rm nongd}=\dfrac{(2\epsilon-1)(1-3M/r_{c})}{r_{c}\sqrt{4 \epsilon (1-\epsilon)}}.
\end{equation}
The variation of $\omega _{\rm nongd}$ with radial distance for different choices of $\epsilon$ has been depicted in \ref{Fig_02b}. As evident from the figure the precession frequency $\omega _{\rm nongd}$ vanishes at the photon circular orbit located at $r_{\rm ph}=3M$ and then remains non-zero throughout, which is in direct agreement with \ref{Eq_dS_ng}. As \ref{Eq_dS_ng} further reveals, for $\epsilon <0.5$ the spin frequency is positive, while for $\epsilon >0.5$ it is negative. This feature can also be observed from \ref{Fig_02b} as well. Finally, the precession frequency of a gyroscopic moving along a circular non-geodesic trajectory becomes,
\begin{equation}
\mathcal{G}_{\rm nongd}=2\pi\left(1-\frac{\omega_{\rm nongd}}{\Omega_{\rm nongd}}\right)
=2\pi\left(1-\dfrac{1-(3M/r_{c})}{\sqrt{4\epsilon \left(1-2M/r_{c}-\dfrac{\Lambda r_{c}^2}{3}\right)\left(1-\epsilon\right)}}\right)
\end{equation}
It is clear that $\mathcal{G}_{\rm nongd}$ becomes $2\pi$ on the photon circular orbit, since $\omega _{\rm nongd}$ vanishes there. Moreover if one expands the expression for $\mathcal{G}_{\rm nongd}$ to leading order in $\Lambda$, then the effect will be smaller compared to that for Schwarzschild. Since for geodesic observers, $\mathcal{G}_{g}$ is identical to that in Schwarzschild spacetime, the above can act as a discriminator between geodesic and non-geodesic observer in terms of spin frequency.
\subsection{Spin precession in asymptotically flat charged Galileon black hole} \label{subsec:asmp_flat}

\begin{figure*}
\begin{center}

\includegraphics[scale=0.5]{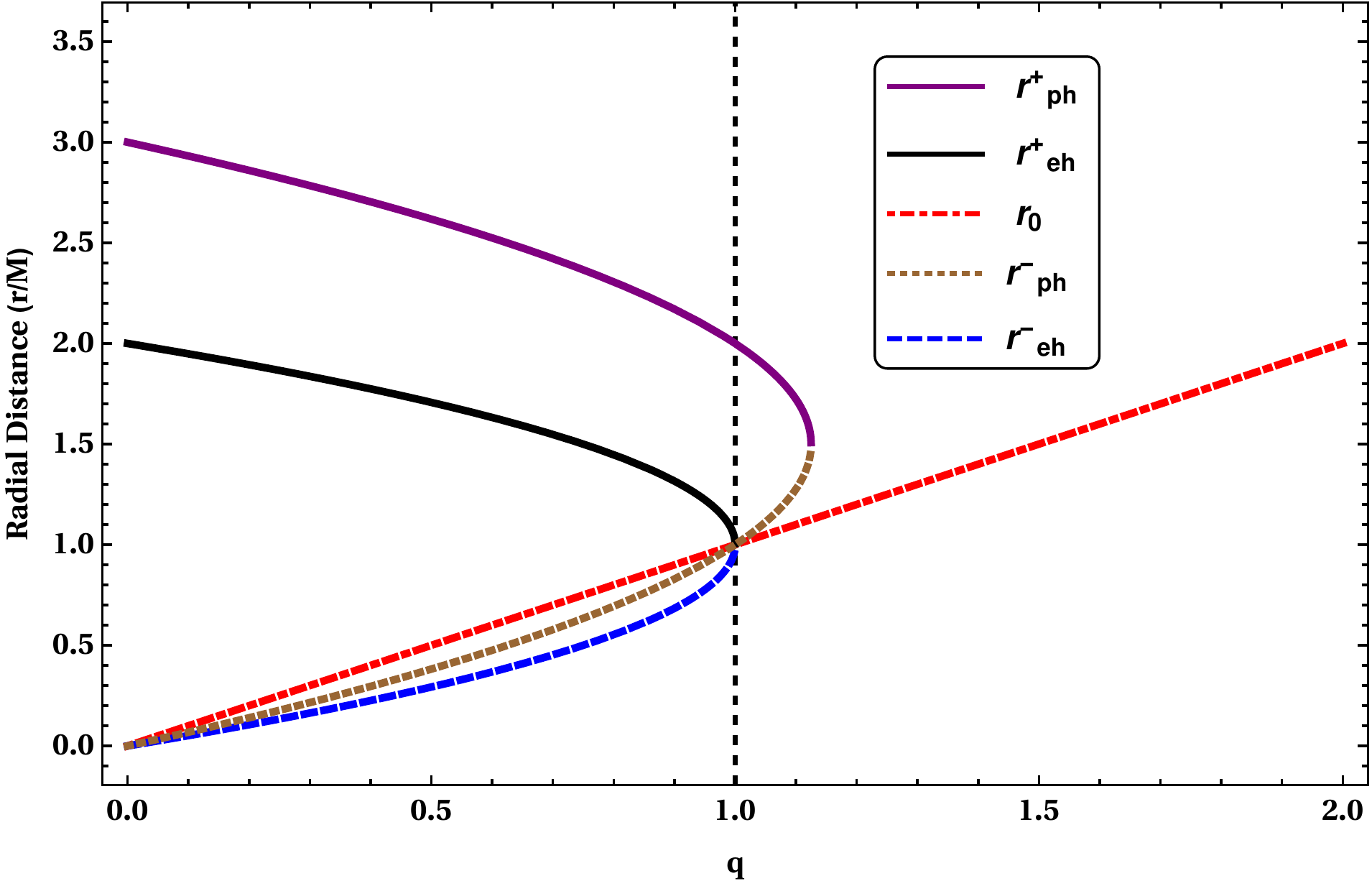}
\caption{The above figure presents the horizon structure in the black hole spacetime associated with Horndeski theories for positive values of $q$. The event horizon $r_{\rm eh}^{+}$ (thick black line) is always at a greater radius compared to $r_{\rm eh}^{-}$ (blue, dotted line), while they coincide at the extremal limit (i.e., $q=1$). The photon radius $r_{\rm ph}^{+}$ (thick, violet line) is always the outermost one, while $r_{\rm ph}^{-}$ (brown, heavily dotted curve) is within the outer event horizon. The radius $r_{0}$ (red, dot dashed line) is always within the outer photon radius and is only an observable for $q>9/8$, when the photon orbits become non-existent.}
\label{Fig_04}
\end{center}
\end{figure*}

\begin{figure*}
\begin{center}

\includegraphics[scale=0.6]{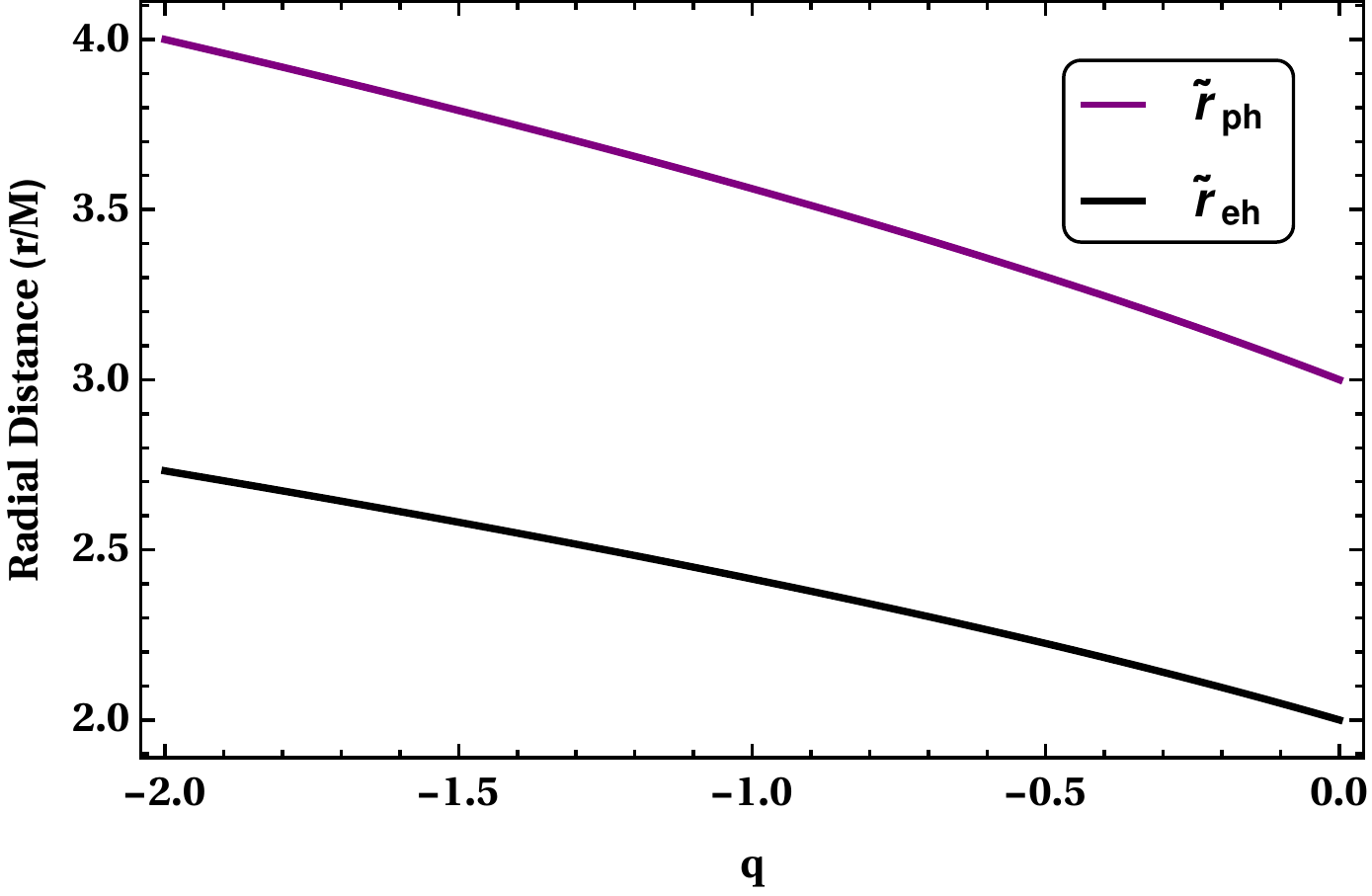}
\caption{The above figure depicts the horizon structure in the asymptotically flat black hole spacetime in charged Galileon theories for negative values of $q$. There is a single event horizon $\tilde{r}_{\rm eh}$ (thick black line) and a photon orbit $\tilde{r}_{\rm ph}$ (thick violet line). The photon orbit radius is always greater then the event horizon. For negative $q$ no such radius $r_{0}$ exists where $\omega _{\rm g}$ vanishes.}
\label{Fig_05}
\end{center}
\end{figure*}

Having discussed the spin precession for both geodesic and non-geodesic observers in the simpler context of Schwarzschild de-Sitter spacetime, let us now concentrate on the asymptotically flat branch of a charged Galileon black hole. As elaborated earlier, this corresponds to an exact solution in the context of Horndeski theories, in which a scalar field couples non-minimally to gravity and a gauge field. In this case as well the metric elements of the four-dimensional spherically symmetric spacetime, following \ref{charge_Gal_BH} can be written as in \ref{eq:Metric} with, $e^{\nu}=1-(2M/r)+(M^2 q/r^2)=e^{-\lambda}$, where $M^{2}q=\gamma(Q^{2}+P^{2})/4\beta$ \cite{Babichev:2015rva}. Note that $q$ can take negative values as well, which will be a characteristic property of these Horndeski theories. Moreover, for positive values of $q$, there will be two horizons located at $r_{\rm eh}^{\pm}=M\pm M\sqrt{1-q}$ (as depicted in \ref{Fig_04}) respectively.
Thus for $q>1$, there will be a naked singularity. While for negative values of $q$, there will be no naked singularity, but only a single horizon located at $\tilde{r}_{\rm eh}=M+M\sqrt{1+|q|}$ (see \ref{Fig_05}). In what follows we will treat the positive and negative values of $q$ separately. Given the spacetime structure one can immediately compute the angular velocity $\Omega _{\rm g}$ of a gyroscope moving in a circular geodesic as well as its spin frequency $\omega _{g}$ using \ref{Ang_Vel_Geo} and \ref{spin_rot} respectively. This yields,
\begin{equation}\label{spin_MD}
\omega_{\rm g}=\Omega_{\rm g}\sqrt{1-\dfrac{3M}{r_{c}}+\dfrac{2 M^2 q}{r_{c}^2}}
=\sqrt{\left(\dfrac{M}{r_{c}^3}-\dfrac{M^2 q}{r_{c}^4}\right)\left(1-\dfrac{3M}{r_{c}}+\dfrac{2 M^2 q}{r_{c}^2}\right)}~.
\end{equation}
From the above expression for $\omega _{g}$ it is clear that the spin frequency vanishes at, $r_{c}=r_{0}=qM$ and at the outer and inner photon circular orbits located at $r_{c}=r_{\rm ph}^{\pm}=(3M/2)\{1\pm\sqrt{1-(8/9)q}\}$ respectively, for positive values of $q$ (see \ref{Fig_04}). On the other hand, for negative values of $q$, the spin frequency will vanish only on the photon circular orbit, located at $\tilde{r}_{\rm ph}=(3M/2)\{1+\sqrt{1+(8/9)|q|}\}$ (illustrated in \ref{Fig_05}). 

\begin{figure}
\subfloat[The above figure depicts the variation of spin frequency ($\omega_{\rm g}$) with radial distance when the parameter $q \leq 1$. Even though in this case $\omega _{\rm g}$ vanishes at both the photon orbits and at $r_{0}$, since the inner photon orbit ($r_{\rm ph}^{-}$) as well as $r_{0}$ lies within the event horizon ($r_{\rm eh}^+$), they are not accessible to an outside observer. It is clear that $\omega_{\rm g}$ has a maxima at $r_c=3 M$ for $q=1$, and the maxima gradually shifts away from $3M$ with a decrease in the $q$ value. \label{Fig_06a}]{\includegraphics[height=5.5cm,width=.49\linewidth]{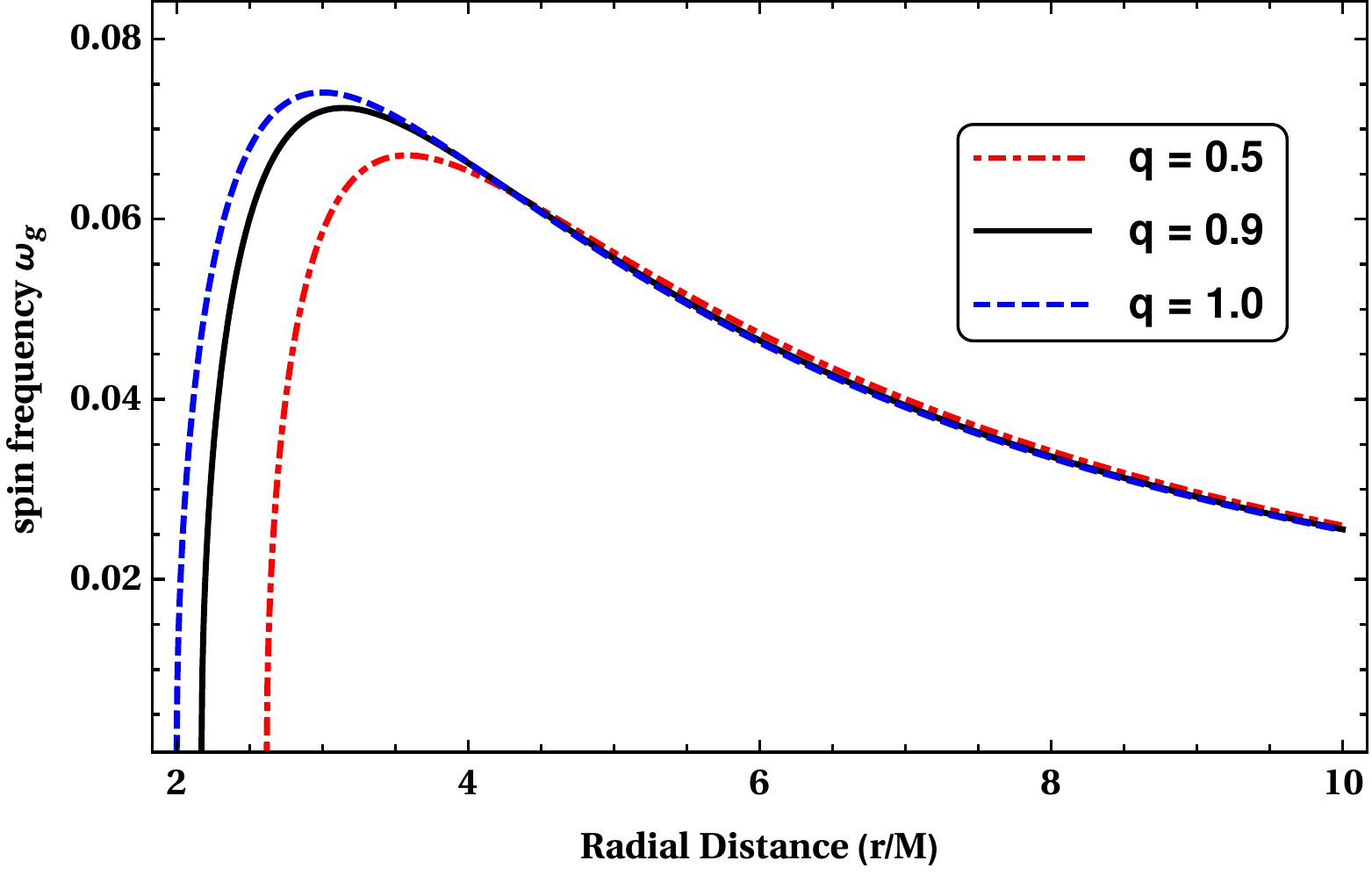}}
\hfill
\subfloat[The above figure depicts how the spin frequency $\omega _{\rm g}$ varies with radial distance, when $1\leq q<9/8$. For $q=1$, the spin frequency $\omega _{\rm g}$ vanishes at the outer photon orbit $r_{\rm ph}^{+}$ alone (as shown by the red dot dashed curve). For other values of $q$, larger than unity but less than $9/8$, the spin frequency $\omega _{\rm g}$ vanishes at three distinct radii, at the outer photon orbit, the inner photon orbit and at $r_{0}$ respectively. Note that $\omega _{\rm g}$ can not have any real value within $r_{\rm ph}^{\pm}$.\label{Fig_06b}]
{\includegraphics[height=5.5cm,width=.49\linewidth]{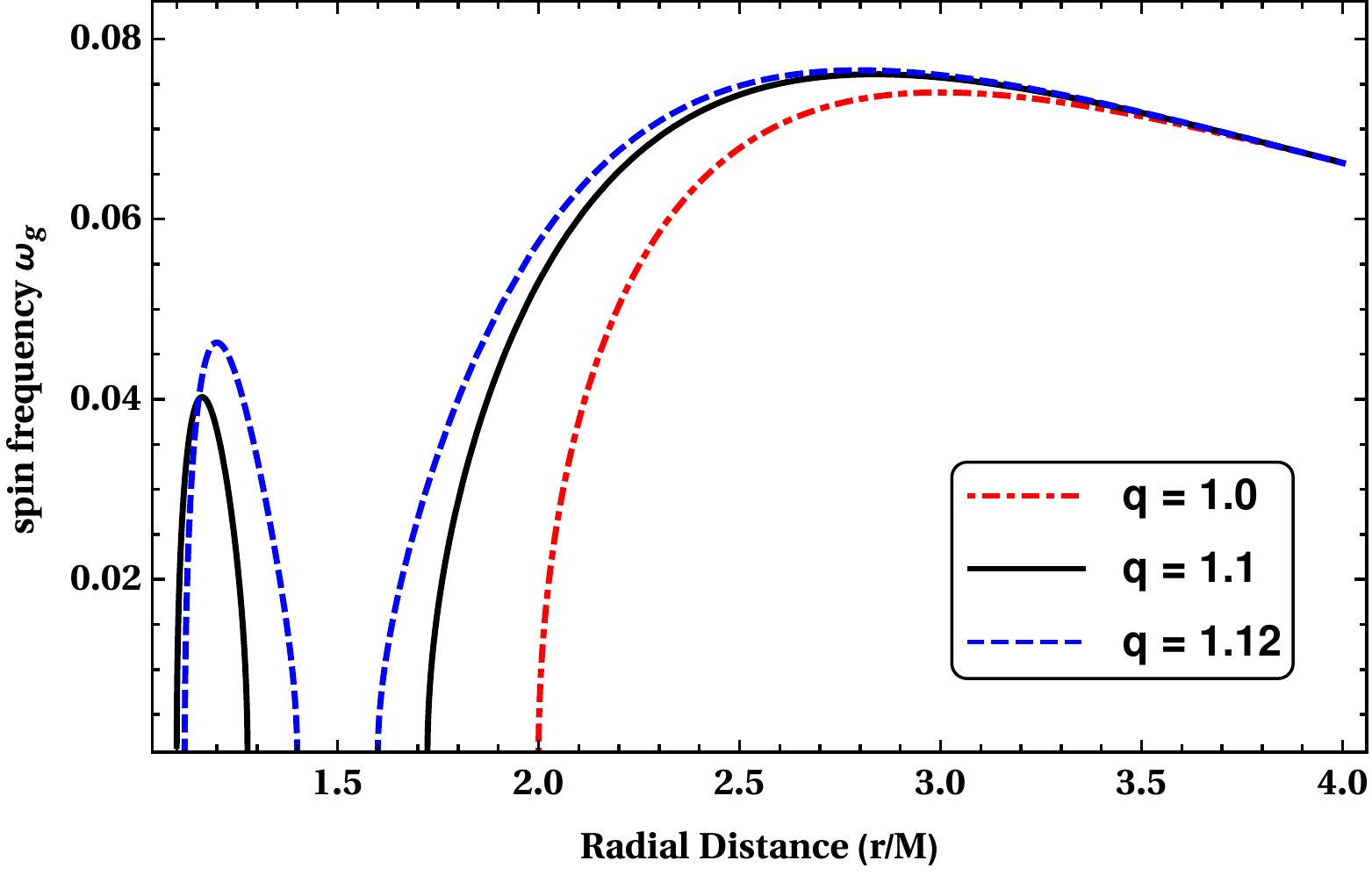}}
\\
\begin{center}
\subfloat[We have illustrated the spin frequency $\omega _{g}$ for three different values of $q$. The first one lies within the range $1<q<9/8$ and hence vanishes at three points (see the red dotted curve). From the right, they are the outer photon orbit, inner photon orbit and $r_{0}$. However for $q>9/8$ (the black and the blue dashed curves) $\omega _{\rm g}$ vanishes \emph{only} at $r_{0}$, which becomes an observable as neither the event horizon nor the photon circular orbit exists in this case.\label{Fig_06c}]{\includegraphics[height=5.5cm,width=0.60\linewidth]{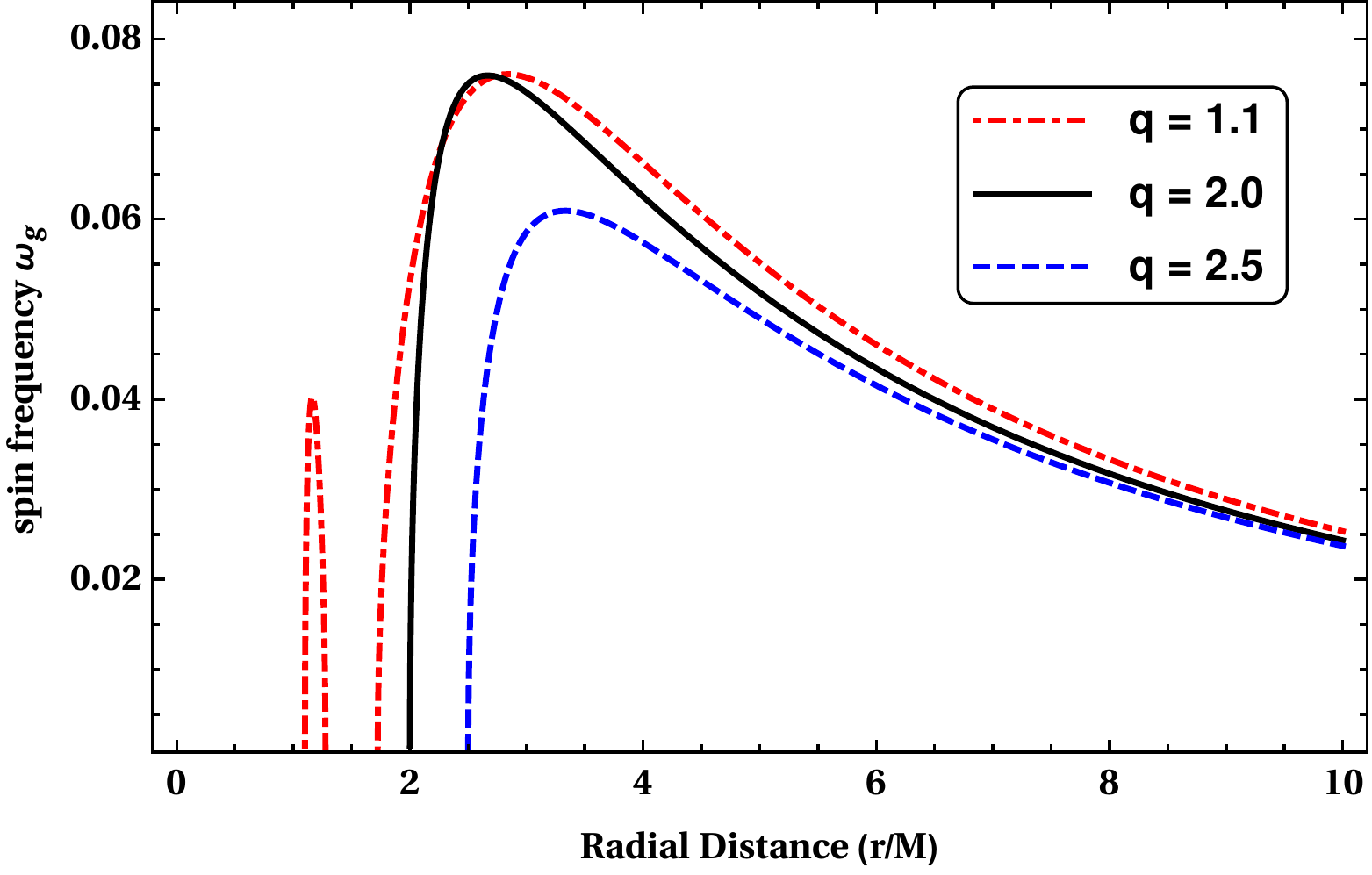}}
\end{center}
\caption{Geodetic spin frequency is being presented for asymptotically flat branch of the charged Galileon black hole solution in Horndeski theories for different choices of $q$.}
\label{Fig_06}
\end{figure}

Given the above spacetime structure there can be two independent situations corresponding to positive $q$ values and negative $q$ values respectively. Each of them can be further sub-divided depending on the behaviour of $\omega _{\rm g}$ for different values of $q$. For positive $q$ values we can have three separate situations:
\begin{itemize}

\item In this case $q<1$. Thus both the horizons as well as the photon circular orbits exist. The spin frequency $\omega _{g}$ vanishes at the photon circular orbits located at $r_{\rm ph}^{\pm}$ as well as at $r_{c}=r_{0}$. But since $r_{0}<r_{\rm eh}^{+}<r_{\rm ph}^{+}$ (see \ref{Fig_04}), it immediately follows that $r_{0}$ will be cloaked by the event horizon and hence it will not be an observable. Thus in this case for an outside observer the spin frequency of a gyroscope moving in a circular geodesic only vanishes at the photon circular orbit as expected. This is illustrated in \ref{Fig_06a}.

\item Another situation corresponds to $1<q<9/8$. In this case a naked singularity appears due to disappearance of the event horizon. However both the photon circular orbits are still in place. In this case $r_{0}<r_{\rm ph}^{-}<r_{\rm ph}^{+}$ (as evident from \ref{Fig_04}) and hence the radius $r_{0}$ is again not accessible, as the spin frequency cannot exist in the region $r_{\rm ph}^{-}<r_{c}<r_{\rm ph}^{+}$. This is depicted in \ref{Fig_06b}.

\item Finally for $q>9/8$, neither the event horizon nor the photon circular orbit exist and hence the spin  frequency $\omega _{\rm g}$ vanishes \emph{only} on the radius $r_{0}=qM$. Hence the radius $r_{0}$ will become an observable. Thus in this particular case one can have a gyroscope moving in a circular orbit with \emph{zero} angular velocity (i.e., it will not move but will remain stationary) and as a consequence $\omega _{\rm g}=0$ and hence the spin vector will also not change direction. This is presented in \ref{Fig_06c}.
 
\end{itemize}
Thus if it is possible to observe the spin frequency $\omega _{\rm g}$ to be vanishing at some finite radius, then it is possible to argue about existence of a naked singularity in this spacetime. This generalizes the claim of \cite{Chakraborty:2016mhx} for Horndeski theories as well. Thus motion of a gyroscope can indeed discriminate naked singularity from event horizon. On the other hand, for negative $q$ values the event horizon is omnipresent as evident from \ref{Fig_05} and further there exist no such radius $r_{0}$, where the spin frequency $\omega _{\rm g}$ vanishes. Thus in this case one will have the usual behaviour for spin frequency $\omega _{\rm g}$, e.g., vanishing at infinity and at photon circular orbit,  having no such non-trivial features. 

After addressing the issue of spin frequency in the context of geodesic motion, let us spell out the expression for geodetic precession frequency  
\begin{equation}
\mathcal{G}_{\rm g}=2 \pi \left(1-\sqrt{1-\dfrac{3M}{r_{c}}+\dfrac{2 M^2 q}{r_{c}^2}}\right)~,
\end{equation}
which can be obtained by using \ref{Geod_prec} along with the expression for the metric elements. It turns out that the precession frequency vanishes on the photon circular orbits ($r_{\rm ph}^{\pm}$) but remains finite for $r_{c}>r_{\rm ph}^{+}$ (also evident from \ref{Fig_06a}). Further it is possible to compute the difference between geodetic precession frequencies of the asymptotic branch of charged Galileon black hole and Schwarzschild solution, which turns out to be negative for positive $q$ and vice versa.  

\begin{figure}[htp]
\subfloat[The above figure shows variation of $\omega_{\rm nongd}$ with radial distance, while keeping $\epsilon$ fixed at $0.7$. The spin frequency vanishes on the photon orbits while a minima appear at $r_c=3M(1-\sqrt{1-2q/3})$ for $q<3/2$. For $q>9/8$, the photon orbits no more exist and the precession is nonzero everywhere. The existence of minima can be used to probe the existence of naked singularity.\label{Fig_07a}]{\includegraphics[height=5.5cm,width=.49\linewidth]{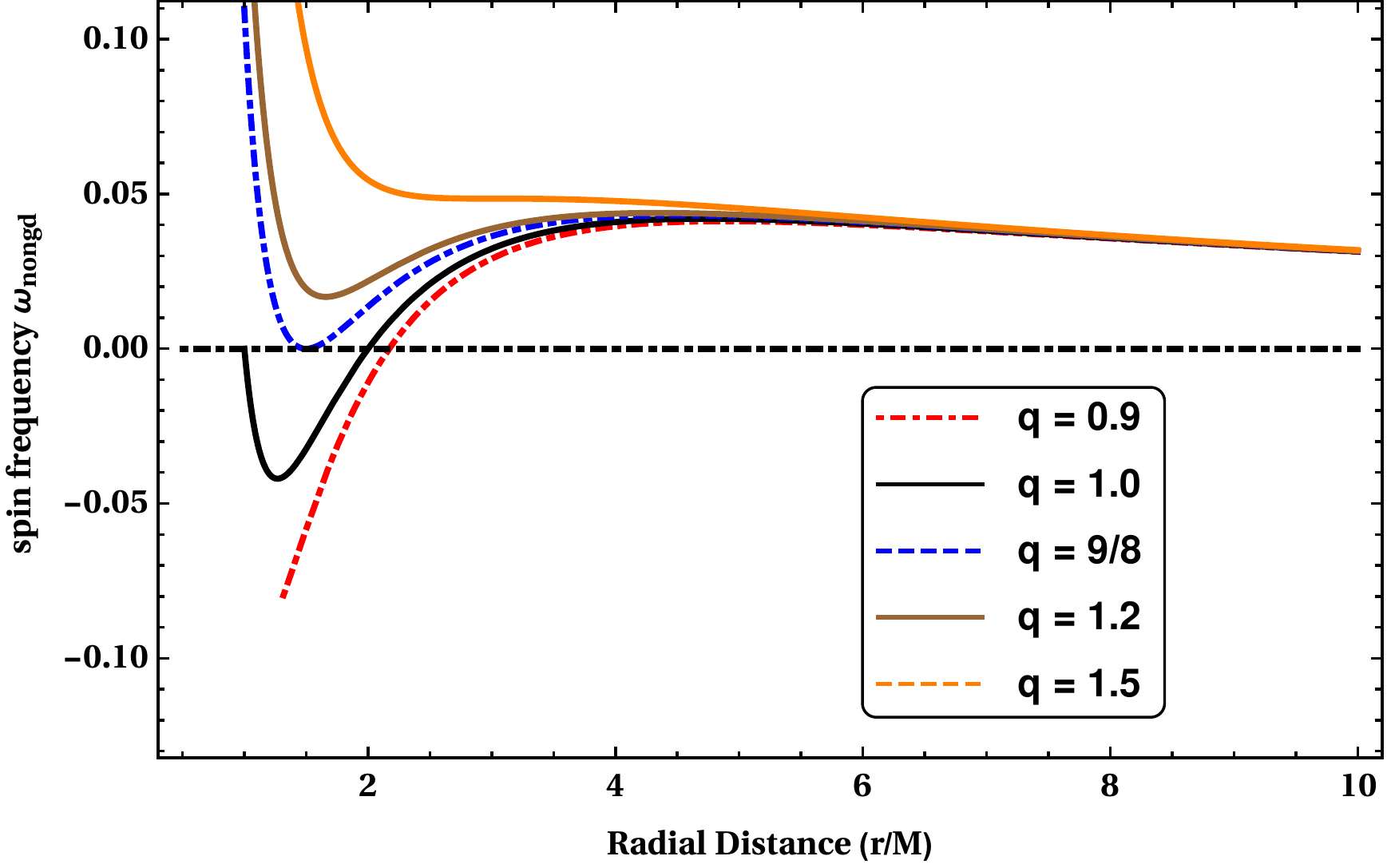}}
\hfill
\subfloat[In the above figure we took $q=0.9$ and have plotted the variation of $\omega_{\rm nongd}$ with radial distance for different values of $\epsilon$. The precession vanishes as $r_{c}$ goes to infinity and it reaches smaller and smaller values as $\epsilon$ becomes close to $0.5$. For $\epsilon >0.5$ the spin frequency $\omega _{\rm nongd}$ is positive while for $\epsilon<0.5$ it is negative.\label{Fig_07b}]{\includegraphics[height=5.5cm,width=.49\linewidth]{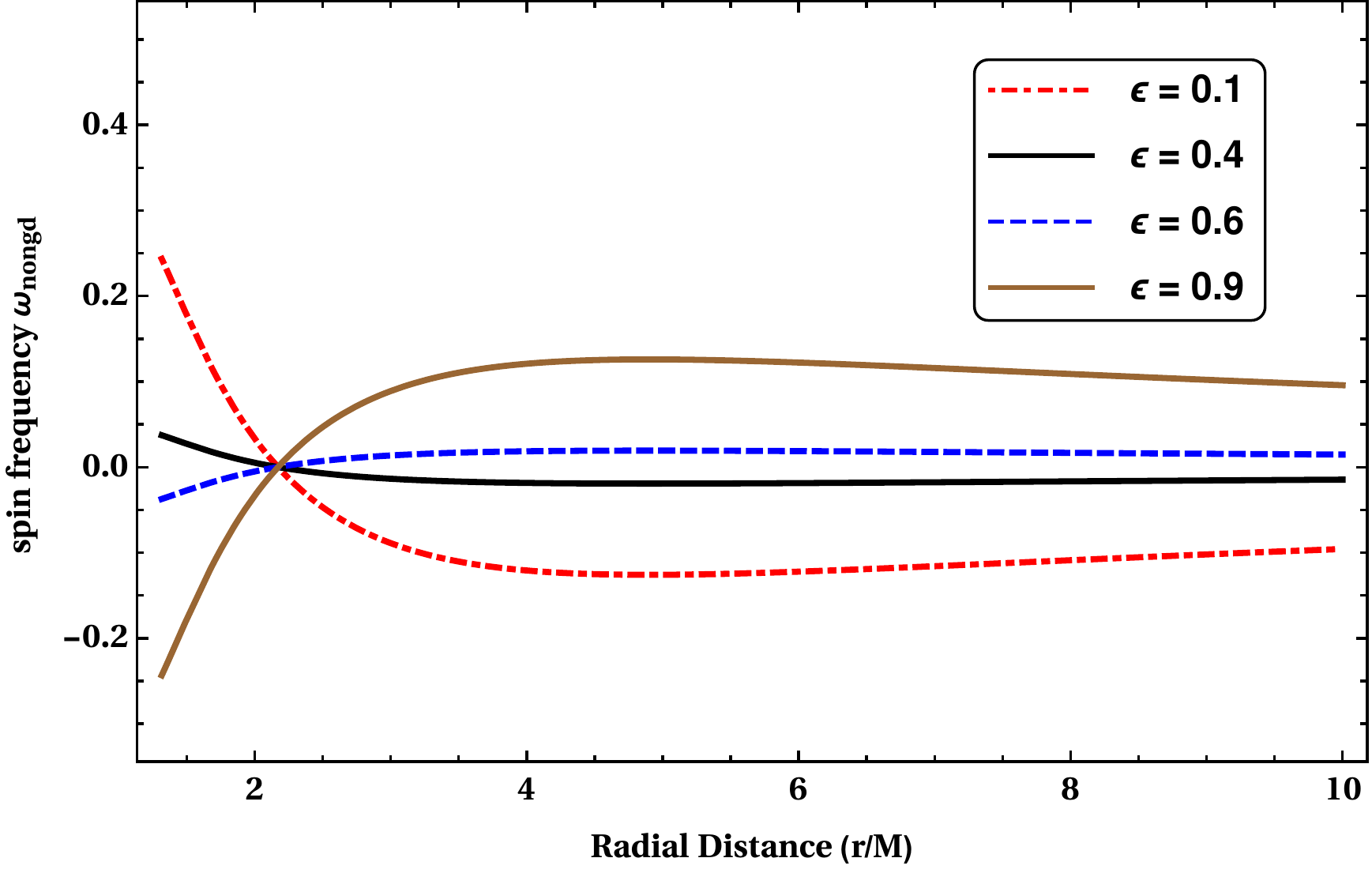}}
\\
\subfloat[The above figure depicts the spin frequency $\omega _{\rm nongd}$ as a function of radial distance $r_c$ and charge parameter $q$ for non-geodesic observers with $\epsilon=0.3$. The contour representing $\omega _{\rm nongd}=0$ has also been presented. \label{Fig_07c}]{\includegraphics[height=8.5cm,width=.49\linewidth]{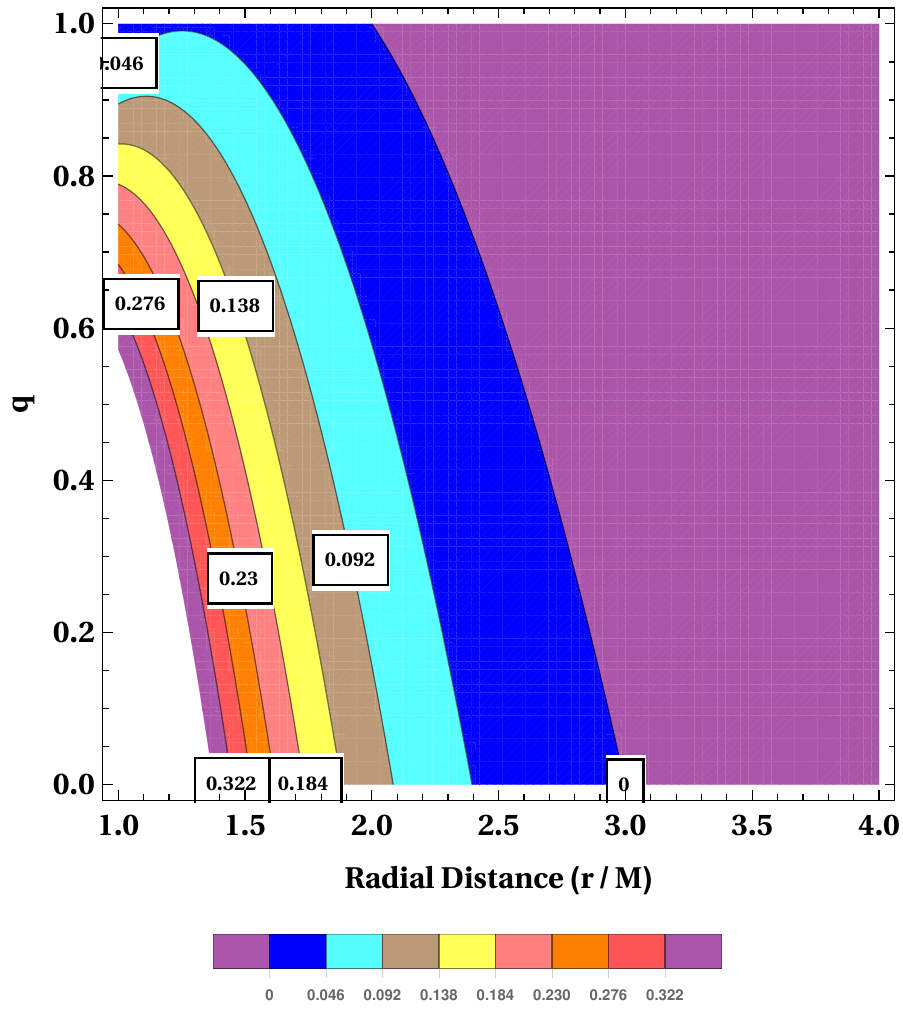}}
\hfill
\subfloat[The above figure illustrates the variation of the spin frequency $\omega _{\rm nongd}$ with the radial distance $r_c$ and the charge parameter $q$ for non-geodesic observers with $\epsilon=0.7$. The contour with $\omega _{\rm nongd}=0$ has also been shown.\label{Fig_07d}]{\includegraphics[height=8.5cm,width=.49\linewidth]{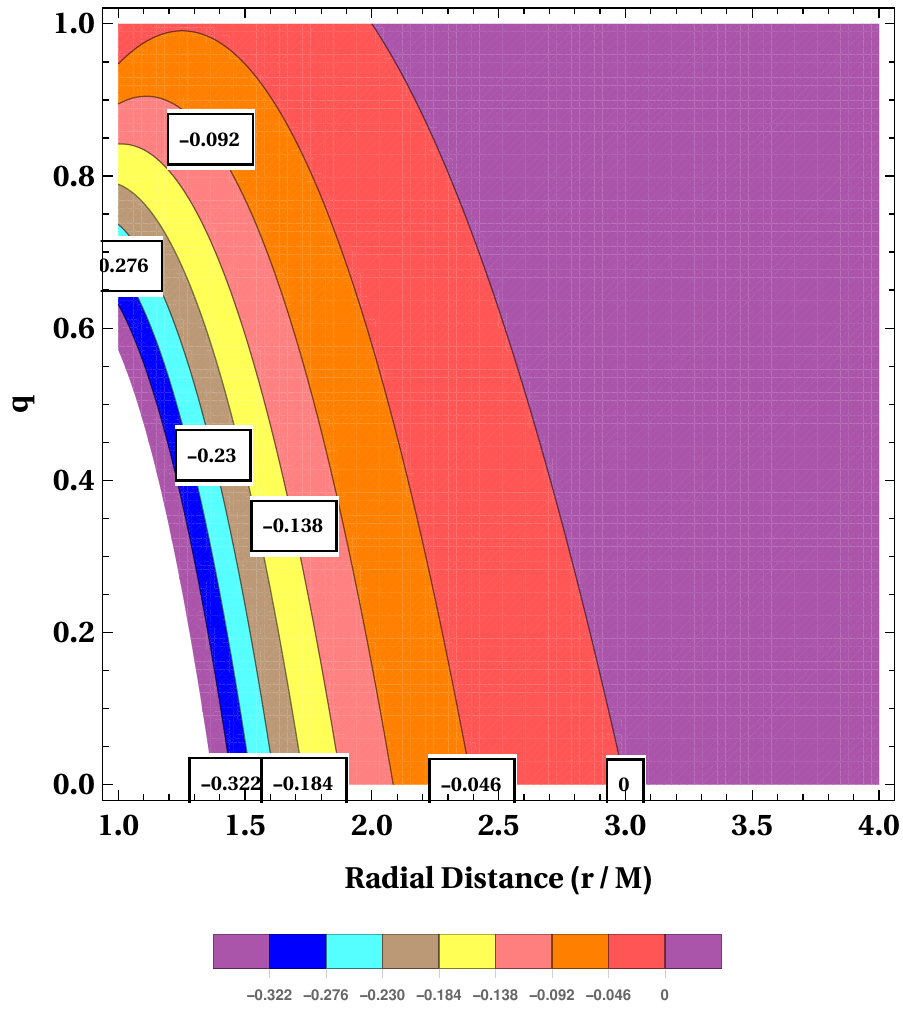}}
\caption{Spin frequency is being illustrated for non-geodesic trajectories in the context of asymptotically flat black hole solutions associated with charged Galileon theories.}
\label{Fig_07}
\end{figure}

On the other hand, for non-geodesic trajectories the spin frequency of the gyroscope can be obtained using \ref{FD_Rot}, such that,
\begin{eqnarray}
\frac{\omega_{\rm nongd}}{\Omega _{\rm nongd}}&=&\dfrac{1-(3M/r_{c})+(2M^{2}q/r_{c}^2)}{\sqrt{4\epsilon(1-\epsilon)\{1-(2M/r_{c})+(M^{2}q/r_{c}^2)\}}}~. \\
\text{or}, \qquad  \omega_{\rm nongd} &=& \dfrac{\left(2\epsilon-1 \right)\{1-(3M/r_{c})+(2M^{2}q/r_{c}^2)\}}{r_{c}\sqrt{4\epsilon(1-\epsilon)}}~.
\label{Eq_MD_nongd}
\end{eqnarray}
As evident from the previous discussion, the parameter $\epsilon$ can neither be zero nor unity, since these values for $\epsilon$ will render the above discussion invalid. Moreover alike the case for geodesic observer, here also we can have $\omega_{\rm nongd}=0$ (see \ref{Fig_07a} and \ref{Fig_07c}). But they are situated precisely at the photon orbit ($r_{\rm ph}^{\pm}$) for any nonzero value of $\epsilon$ (except for $\epsilon$ being $0$ or $1$) and with $\epsilon=0.5$, $\Omega_{\rm nongd}$ become identically zero for any value of $r_{c}$. Further the spin frequency exhibits a minima at $r_{c}=3 M (1-\sqrt{1-2q/3})$, which is within the photon circular orbit. While for $q>9/8$ the photon circular orbit disappears and the minima becomes an observable. This provides yet another root to probe existence of naked singularity, using non-geodesic observers. While for $\epsilon>0.5$, the spin frequency is always positive, but for $\epsilon<0.5$ it is negative, in conformity with \ref{Eq_MD_nongd} (see also \ref{Fig_07b} 
and \ref{Fig_07d}). Finally the precession frequency for the non-geodesic observer becomes,
\begin{equation}
\mathcal{G}_{\rm nongd}=2 \pi \left(1-\dfrac{1-(3M/r_{c})+(2M^{2}q/r_{c}^2)}{\sqrt{4\epsilon(1-\epsilon)\{1-(2M/r_{c})+(M^{2}q/r_{c}^2)\}}}\right)~.
\end{equation}
Similar to the previous case, for this solution as well we can expand the above expression around $q=0$ and hence compute the deviation from \gr, which turns out to be negative for positive $q$ and vice versa. 

Let us now impose the corresponding bound on the charge parameter $q$ using the \GPB\ experiment. For that purpose we bring in the Newton's constant, while keeping the speed of light at unity.  In \GPB\ experiment one considers a satellite orbiting earth carrying gyroscopes. Since the experiment was carried out for a gyroscope moving on a geodesic, the precession $\mathcal{G}_{\rm g}$ will be applicable here. The number of complete revolutions of such a satellite per day is, $n={(24 \times 60)}/{97.65} \approx 15$. Thus in one year total number of revolutions would correspond to $365 \times n$. Hence the precession per revolution must be within the following range: $(-4.199 \times 10^{-6},~2.601 \times 10^{-6})$ arc-sec. Applying this result to the spin precession in the context of asymptotically flat branch of the charged Galileon black hole we obtain $|q|<0.03$. The above bound on the charge parameter is consistent with other solar system tests, e.g., perihelion precession of Mercury and 
bending angle of light. In particular, it turns out that the constraint from perihelion precession of Mercury corresponds to $|q|<0.024$ and hence is stronger than the above bound from \GPB\ \cite{Bhattacharya:2016naa}. On the other hand, the constraint from \GPB\ is much better than the corresponding bound from bending angle of light, which corresponds to $|q|<0.046$ \cite{Bhattacharya:2016naa}. Thus the constraint from \GPB\ indeed improves the bound from bending angle of light by $\sim 35\%$, however it is the perihelion precession of Mercury, which provides the most stringent bound.

As we have elaborated earlier, this solution can have several origins. For example, an identical spacetime metric appears in the context of Maeda-Dadhich solution, which originates from the Kaluza-Klein reduction of a higher dimensional solution in the context of Einstein-Gauss-Bonnet gravity. Intriguingly, the Maeda-Dadhich solution must have negative $q$ and hence the associated spin frequency will be higher than its general relativistic counterpart. In this physically distinct scenario as well the spacetime is geometrically indistinguishable from the asymptotically flat black hole solution pertaining to charged Galileon theories and hence the results presented in this section can be applied in a straightforward manner. Hence the bounds on the parameter $q$ from \GPB\ will also translate into bounds on the associated parameters in other alternative gravity models. 
\subsection{Spin precession in asymptotically de-Sitter charged Galileon black hole}\label{subsec:asmp_ds}

\begin{figure}[htp]
\subfloat[The horizon structure has been depicted with $\Lambda=10^{-1}M^{-2}$, for different values of $q$. For $q > 1.03897$, the event horizons do not exist and a naked singularity appears. Similarly for $q>9/8$, photon orbits disappear and spin frequency would only vanish at $r_{c}=r^{\pm}_{0}$. Increasing the charge parameter further such that $q>1.46808$, the spin frequency become imaginary at any value of $r$. \label{Fig_12a}]{\includegraphics[height=5.2cm,width=.49\linewidth]{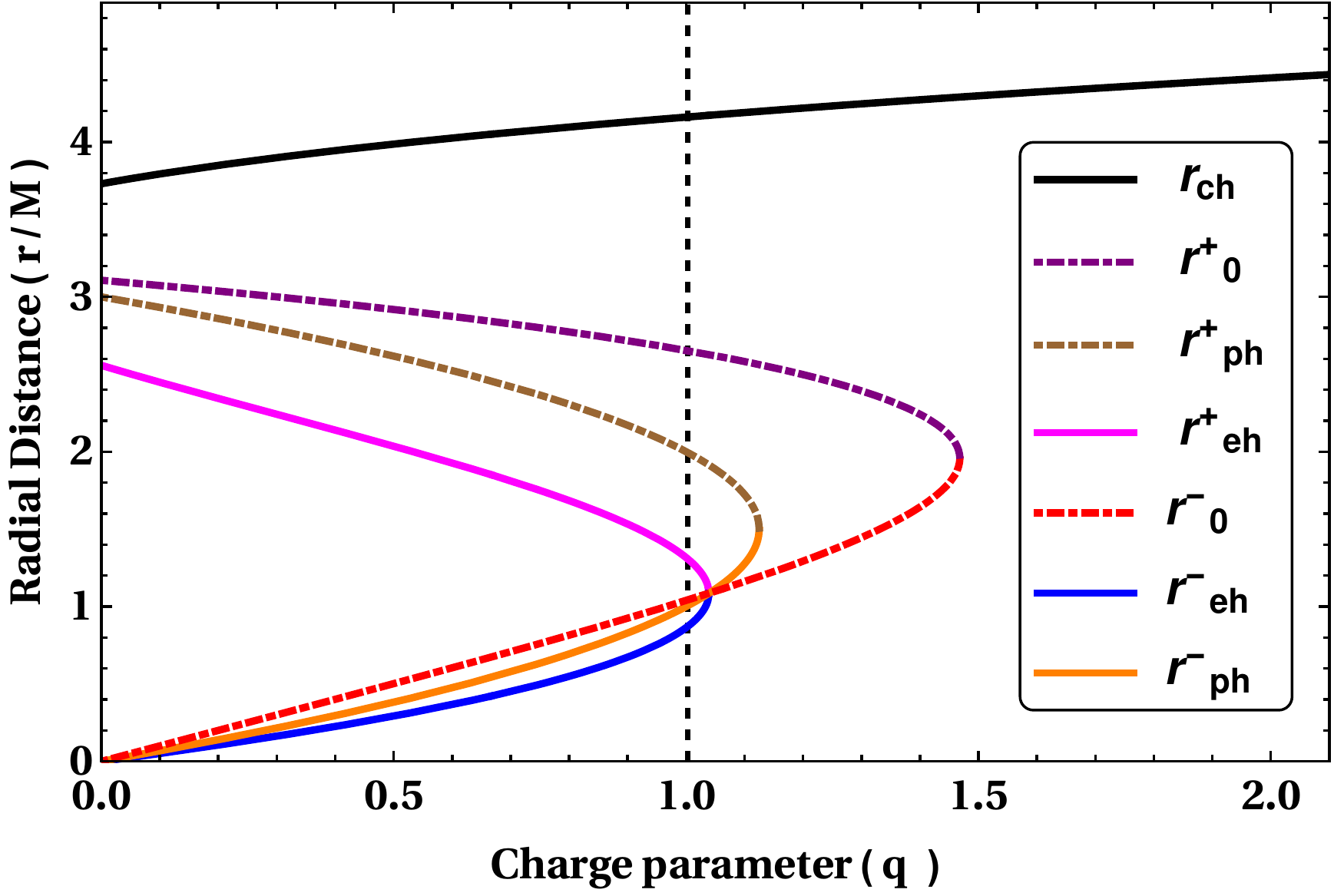}}
\hfill
\subfloat[The variation of horizon structure with the Galileon charge $q$ is being presented for $\Lambda=10^{-2}M^{-2}$. In this case the event horizons disappear for $q \gtrapprox 1.00338$ while for $q>9/8 $ the photon circular orbits cease to exist. In addition, for $q \gtrapprox 3.16287$, $r_{0}^{\pm}$ no longer exists and hence the spin frequency become imaginary for any value of the radial parameter. \label{Fig_12b}]{\includegraphics[height=5.2cm,width=.49\linewidth]{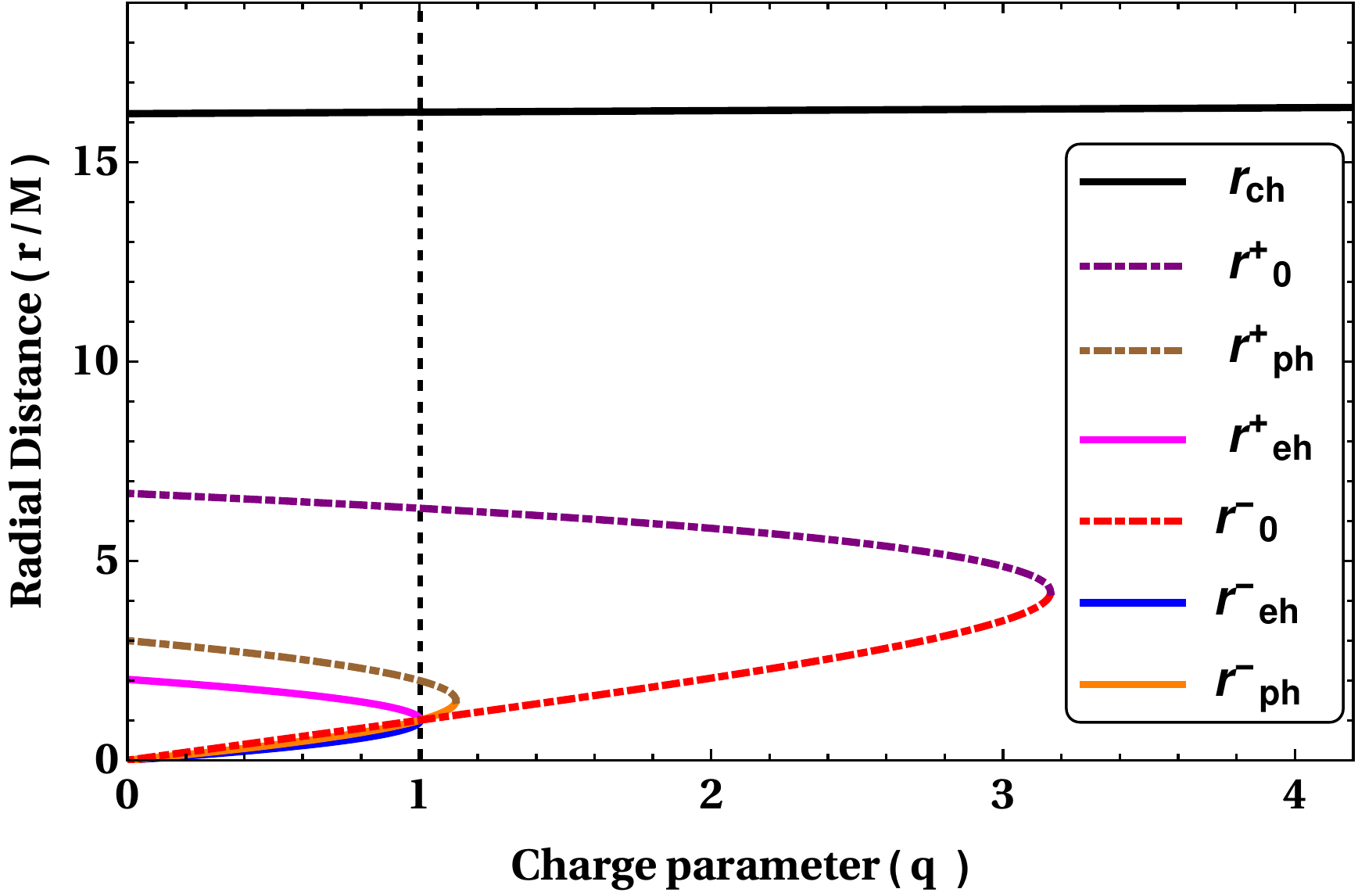}}
\caption{Location of the even horizons, photon orbits and $r^{\pm}_{0}$ are being presented against the charge parameter $q$ for two different choices of the cosmological constant $\Lambda$ in case of an asymptotically de-Sitter charged Galileon black hole.}
\label{Fig_12}
\end{figure}

In this section, we shall consider the asymptotically de-Sitter branch of the charged Galileon black hole within the context of Horndeski theories. The corresponding solution has already been discussed in \ref{Horn_B_Intro} and presented in \ref{dS_charge_BH}. This metric can also be casted in a form similar to \ref{eq:Metric}, such that, $e^{\nu}=e^{-\lambda}=1-(2M/r)-(\Lambda/3)r^{2}+q( M^2/r^{2})$. Here, $\Lambda=|\eta|/\beta$ and $M^2 q=\gamma (Q^{2}+P^{2})/(4\beta)>0$. Given this spherically symmetric solution, the location of the event horizon can be found by solving the algebraic equation
\begin{align}
r^{2}-2Mr-\frac{\Lambda}{3}r^{4}+qM^{2}=0~.
\end{align}
The above equation has three real solutions denoting the cosmological horizon ($r_{\rm ch}$) along with an outer (inner) event horizon located at $r^{\pm}_{\rm eh}$ respectively (see \ref{Fig_12}). While the location of photon circular orbit can be obtained from solving the equation $2=r\nu'$, such that the following algebraic relation can be obtained,
\begin{align}
r^{2}-3Mr+2qM^2=0~.
\end{align}
Thus photon circular orbit is not affected by the presence of an effective cosmological constant and will be located at, $r^{\pm}_{\rm ph}=(3M/2)(1\pm\sqrt{1-8q/9})$ while $\pm$ has their usual meaning of outer and inner photon circular orbits respectively. The angular frequency associated with the motion of a gyroscope is given as,
\begin{align}
\Omega _{\rm g}=\sqrt{\frac{M}{r_{c}^{3}}-\frac{\Lambda}{3}-\frac{q M^2}{r_{c}^{4}}}~.
\end{align}
This would immediately suggest that the spin frequency $\omega_{\rm g}$ becomes
\begin{align}
\omega _{\rm g}=\left(\sqrt{\frac{M}{r_{c}^{3}}-\frac{\Lambda}{3}-\frac{qM^2}{r_{c}^{4}}}\right)\left(1-\frac{3M}{r_{c}}+\frac{2qM^2}{r_{c}^{2}}\right)^{1/2}~.
\label{eq:spin_fre_RN_DS}
\end{align}
Similar to the previous cases here as well, the spin frequency vanishes at the photon circular orbits $r_{\rm ph}^{\pm}$ as well as when $\Omega _{\rm g}$ vanishes. Vanishing of $\Omega _{\rm g}$ corresponds to the presence of additional correction terms to the \EH action and of course has contribution from the cosmological constant. These additional locations where spin frequency of a gyroscope vanishes corresponds to solutions of the following algebraic equation
\begin{align}
\frac{\Lambda}{3}r_{c}^{4}+qM^{2}-Mr_{c}=0
\end{align}
which has two real solutions, denoted by $r_{0}^{\pm}$, while the other two are complex conjugate to each other.

Furthermore, given the structure of the event horizon, photon circular orbit as well as the radius $r^{\pm}_{0}$, illustrated in \ref{Fig_12}, we shall try to understand the behaviour of spin frequency as well as geodetic precession against the charge parameter for a specific value of $\Lambda$ (fixed at $\Lambda=10^{-1}M^{-2}$ for illustration purpose). Keeping this in mind, we will try to present a detailed analysis of the parameter space of the Galileon charge vis-\'{a}-vis the spin frequency of the gyroscope below.

\begin{figure}[htp]
\subfloat[The above figure depicts the variation of $\omega _{\rm g}$ with radial distance for different $q$ values. It is clear that the spin frequency vanishes at the outer photon circular orbit as well as at $r^+_{0}$. The other two radii, namely $r^{-}_{0}$ and the inner photon circular orbit are clocked by the event horizon and hence not visible to an observer. In the case of $q=1$, $r^{+}_{0}$ is located at $\approx 14M$, while the outer photon orbit exactly placed at $2M$. \label{Fig_13a}]{\includegraphics[height=5cm,width=.49\linewidth]{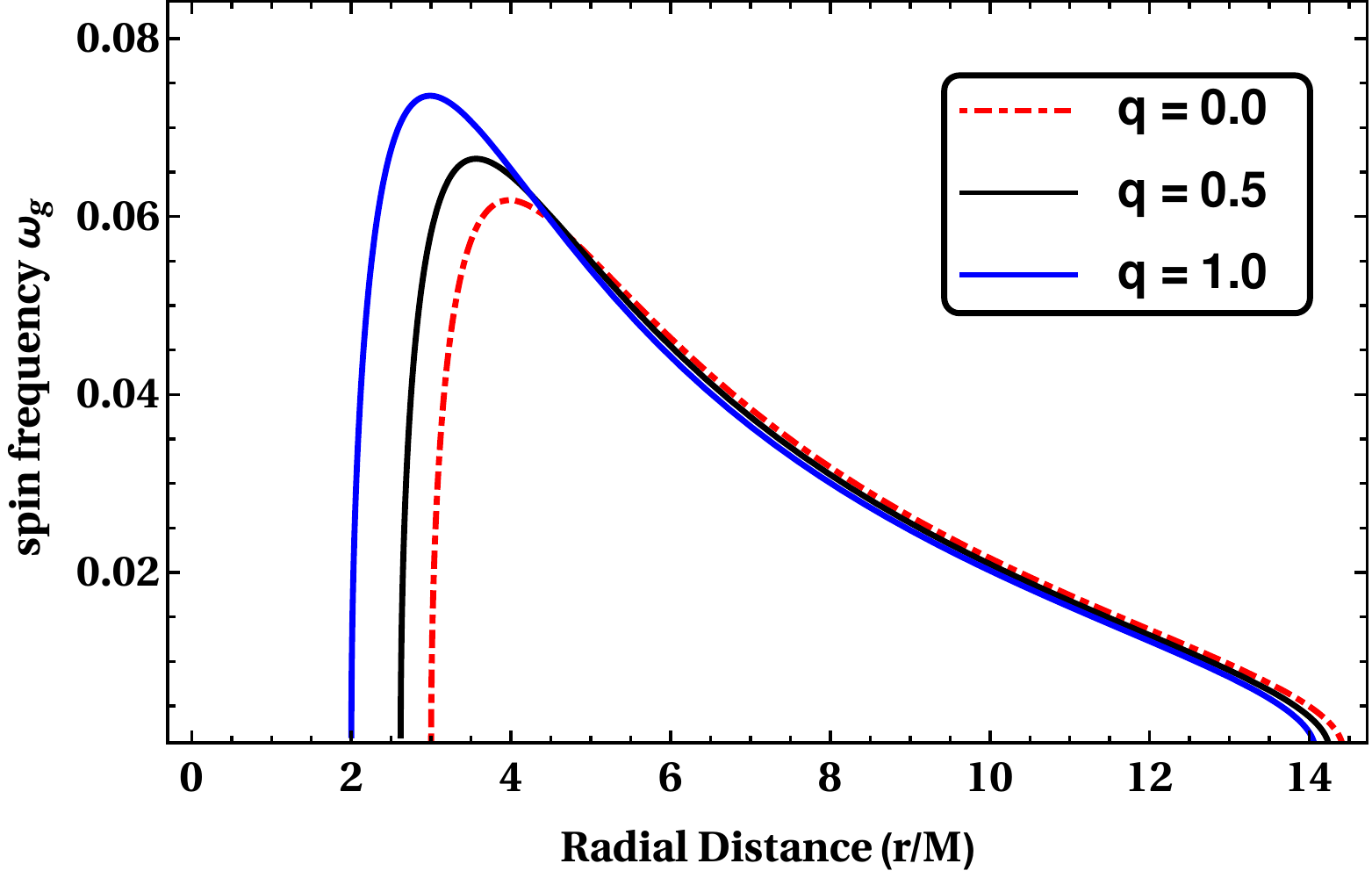}}
\hfill
\subfloat[In case of $q \gtrapprox 1.00033$, there is no event horizon in the spacetime and the singularity is visible. The spin frequency vanishes at $r^{\pm}_0$ as well as at the photon circular orbits. But as the frequency become imaginary within the photon circular orbits, neither the inner photon circular orbit nor $r^{-}_{0}$ would be an observable. For $q>9/8$, no photon orbits exist anymore and $\omega_g$ vanishes only at $r_{c}=r^{\pm}_{0}$.\label{Fig_13b}]{\includegraphics[height=5cm,width=.49\linewidth]{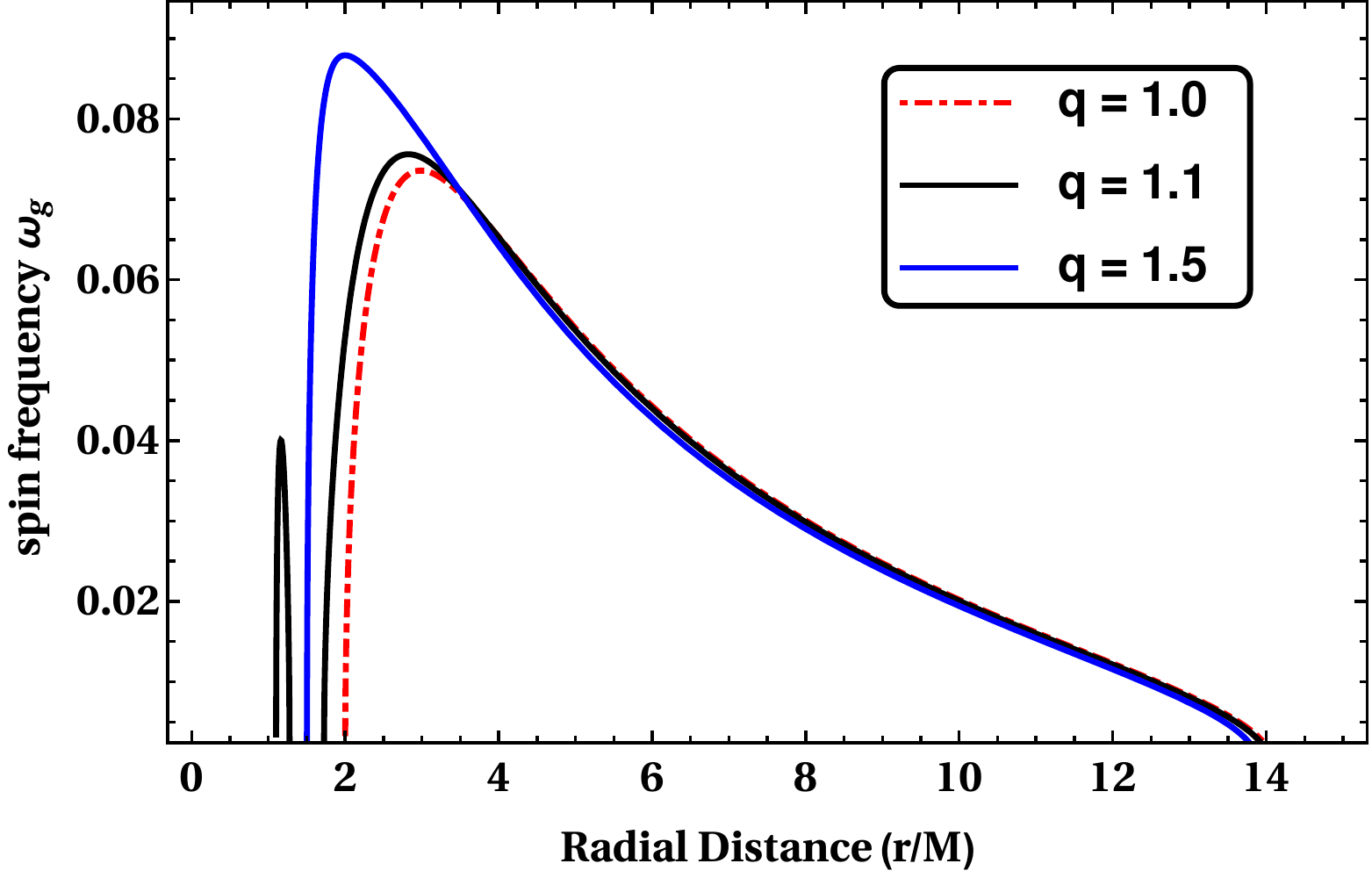}}
\caption{The spin frequency $\omega_g$ is presented in the asymptotically de-Sitter branch of the charged Galileon black hole, where the cosmological constant is being fixed at $\Lambda=10^{-3}M^{-2}$.}
\label{Fig_13}
\end{figure}

\begin{enumerate}

\item For $0<q \lesssim 1.03897 $, both the photon orbits and event horizons exist. Even if the spin frequency $\omega_g$ vanishes at the photon circular orbits as well as at $r_{c}=r^{\pm}_0$, both the inner photon circular orbit and $r^{-}_{0}$ are located inside the outer event horizon as shown in \ref{Fig_12a}, i.e., $\textit{i.e.}$, $r^{-}_{\rm ph}<r^{-}_0 <r^{+}_{\rm eh}$. Hence, only the $r^{+}_{\rm ph}$ and $r^{+}_{0}$ would be noticeable for a distant observer, this is illustrated in \ref{Fig_13a}.
 
\item For $1.03897  \lesssim q \leq 9/8 $, the event horizon does not appear anymore and the singularity would be naked. In this case, as one can easily interpret from \ref{Fig_12a} that $r^{-}_0 < r^{-}_{\rm ph} < r^{+}_{\rm ph}<r^{+}_0$. So, even with a naked singularity, the observable radii where the spin frequency would vanish is $r^+_{0}$ and $r^{+}_{\rm ph}$ respectively. This is because, $\omega_g$ become imaginary within the region $r^{-}_{\rm ph} < r_{c} < r^{+}_{\rm ph}$ and the gyroscope can no longer exists there, see \ref{Fig_13b}.  

\item For $9/8  \leq q \lesssim 1.46808 $, neither the event horizons nor the photon orbits exist. In this case, the spin frequency only vanishes at $r^{\pm}_0$ and unlike the previous situations, both of these would be observable. Finally for $q \gtrapprox 1.46808 $ the spin frequency become imaginary as $\Omega^2_{\rm g}$ become negative. Therefore the gyroscope can no longer exists in the spacetime with the above parameter space. This is presented in \ref{Fig_13b}. 
 
\end{enumerate} 

Following \ref{Geod_prec} and \ref{eq:spin_fre_RN_DS}, it is straightforward to compute the geodetic precession frequency associated with the geometry of the charged Galileon black hole as,
\begin{align}
\mathcal{G}_{\rm g}=2\pi \left(1- \sqrt{1-\frac{3M}{r_{c}}+\frac{2qM^2}{r_{c}^{2}}}\right)
\end{align}
Surprisingly, the above expression is independent of the cosmological constant $\Lambda$ and the constraints on the parameter $q$ would be exactly similar to those presented in \ref{subsec:asmp_flat}. This essentially suggests that the existence of a cosmological constant can not be identified by only studying the geodetic precession frequency. While, the spin frequency $\omega_g$ or in particular, the angular velocity $\Omega_g$ of the gyroscope carry the imprints of the cosmological constant. 

On the other hand, for a gyroscope moving along an accelerated trajectory the associated spin frequency can be written following \ref{Eq:Fermi_dragged} as, 
\begin{align}
\omega _{\rm nongd}=\left(\frac{2\epsilon -1}{r_{c}}\right)\dfrac{1-\frac{3M}{r_{c}}+\frac{2q}{r_{c}^{2}}}{\sqrt{4\epsilon (1-\epsilon)}}
\end{align}
This is in exact agreement with \ref{Eq_MD_nongd} which describes an asymptotically flat spacetime (see \ref{Fig_07} for an elaborate discussion). Following the above expression along with \ref{nongeod_prec}, the non-geodetic precession becomes
\begin{align}
\mathcal{G}_{\rm nongd}=2\pi \left[1-\frac{1-\frac{3M}{r_{c}}+\frac{2q}{r_{c}^{2}}}{\sqrt{4\epsilon (1-\epsilon)}\sqrt{1-\frac{2M}{r_{c}}-\frac{\Lambda}{3}r_{c}^{2}+\frac{q}{r_{c}^{2}}}}\right]
\end{align}
In passing, we would like to point out that the cosmological constant appears in the denominator of the above expression, which is due to the angular velocity $\Omega_{g}$ of the gyroscope. The rest of the properties associated with $\mathcal{G}_{\rm nongd}$ has already been discussed in the previous section and hence will not be repeated here. 
\subsection{Spin precession in Einstein-dilaton-Gauss-Bonnet gravity: The Sotiriou-Zhou solution}

\begin{figure*}[htp]
\begin{center}

\includegraphics[scale=0.6]{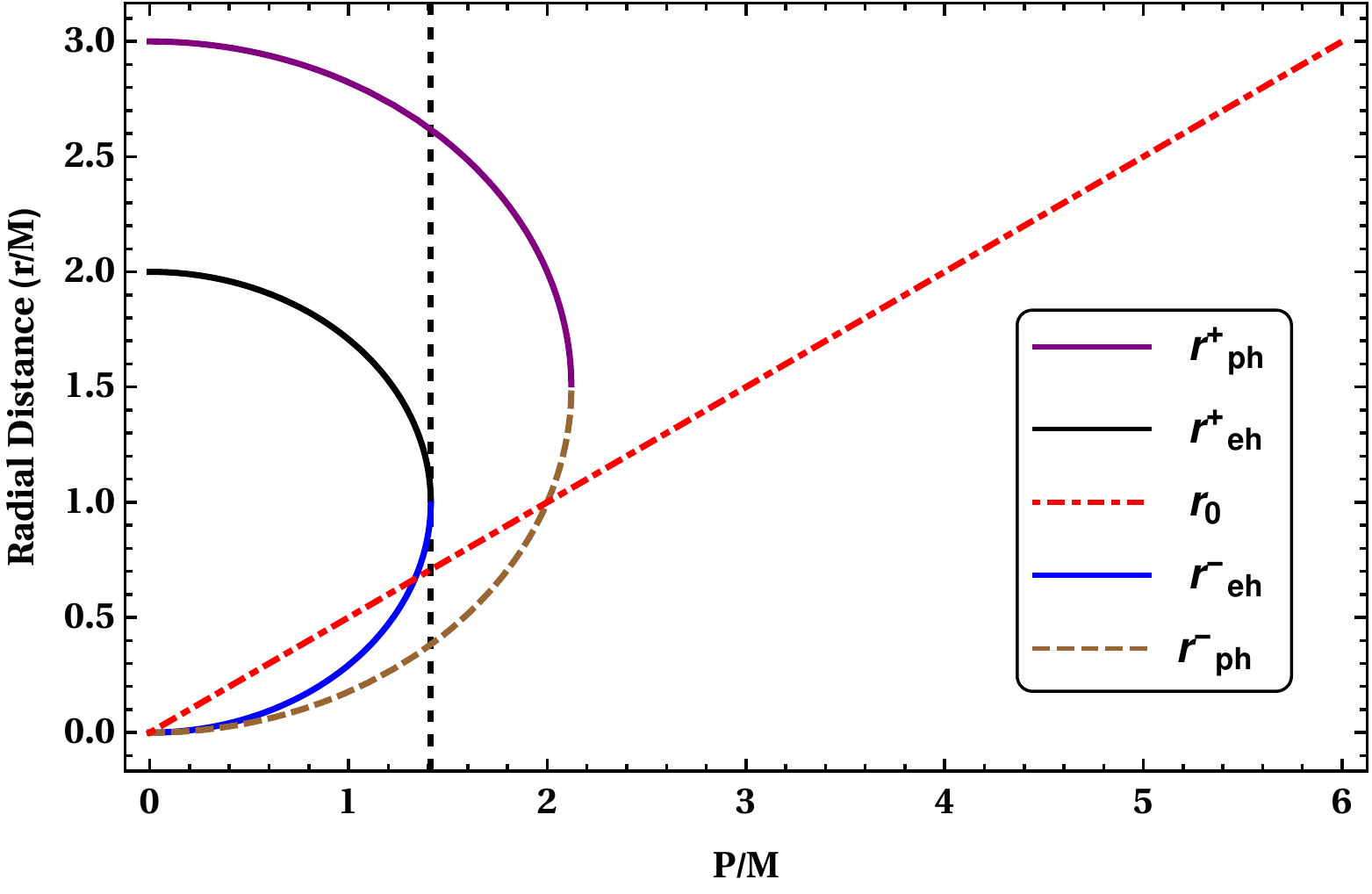}
\caption{The above figure depicts the horizon structure of Sotiriou-Zhou spacetime. The outer event horizon $r_{\rm eh}^{+}$ (thick black line) is always at a greater radius compared to $r_{\rm eh}^{-}$ (blue line), while they coincide at the extremal limit (i.e., $P=\sqrt{2}M$). The photon radius $r_{\rm ph}^{+}$ (thick, violet line) is always the outermost one, while $r_{\rm ph}^{-}$ (brown, dotted curve) is within the inner horizon. The radius $r_{0}$ (red, dot dashed line) is always within the outer photon radius and is only an observable for $(P/M)>3/\sqrt{2}$, when the photon orbits become non-existent.}
\label{Fig_09}
\end{center}
\end{figure*}

In this final section, we will discuss another alternative gravity model and a spherically symmetric solution within its framework. This is again a subclass of Horndeski theories and corresponds to Einstein-dilaton-Gauss-Bonnet gravity. The associated action and the corresponding solution has already been presented in \ref{Horn_B_Intro}. The most interesting fact associated with this solution being, it inherits scalar hair. Thus we hope to put some bounds on the scalar charge using the \GPB\ experiment and discover some interesting features associated with this spacetime as far as spinning object is considered. In this case the two metric components $e^{\nu(r)}$ and $e^{\lambda(r)}$ are different, with the following functional dependences: $e^{\nu}= 1-(2M/r)+(MP^{2}/6r^{3})$ and $e^{\lambda}=1+(2M/r)+\{(8M^{2}-P^{2})/2r^{2}\}$. Note that the metric components are derived from the perturbative solution presented in \ref{eq:SZ_Appx} and for simplicity we have kept only the lowest order term 
presenting the deviation from the Schwarzschild solution. Given the above metric elements, there will be two horizons, whose location can be determined from the algebraic equation $e^{-\lambda}=0$ and corresponds to $r_{\rm eh}^{\pm}=M\pm \sqrt{M^{2}-(P^{2}/2)}$. Thus the location of the event horizon $r_{\rm eh}^{+}$ will always be smaller compared to $2M$ irrespective of the sign of $P$, which is also evident from \ref{Fig_09}. 

\begin{figure}[htp]
\subfloat[The spin frequency $\omega_{\rm g}$ has been plotted against radial distance for three different values of $P/M$. The case $(P/M)=1$ is being presented by the blue, dashed curve vanishing at the outer photon orbit. In the other two cases, $(P/M)<3/\sqrt{2}$ and hence the outer photon orbit always exists on which $\omega _{\rm g}$ vanishes.\label{Fig_10a}]{\includegraphics[height=5cm,width=.49\linewidth]{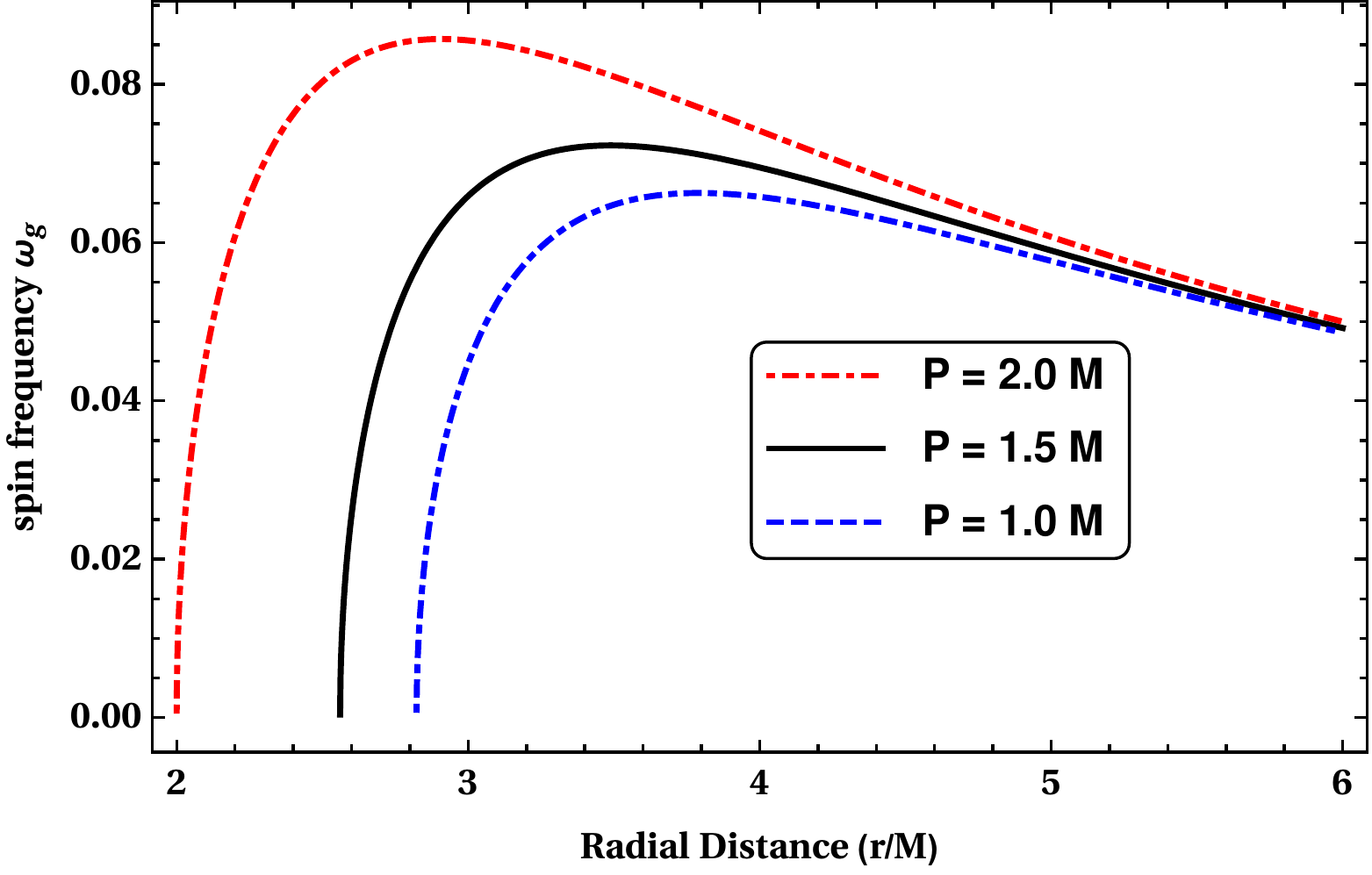}}
\hfill
\subfloat[Spin frequency $\omega _{\rm g}$ is shown for three different values of $P/M$ greater than $\sqrt{2}$. For $\sqrt{2}<(P/M)<3/\sqrt{2}$ (the blue dashed and the thick black curves), the spin frequency vanishes at the outer photon orbit, $r_{0}$ and the inner photon orbit respectively. Beyond this value, i.e., for $P>(3/\sqrt{2})M$, $\omega _{\rm g}$ vanishes only on the radius $r_{0}$ (as the red dot dashed curve depicts).\label{Fig_10b}]{\includegraphics[height=5cm,width=.49\linewidth]{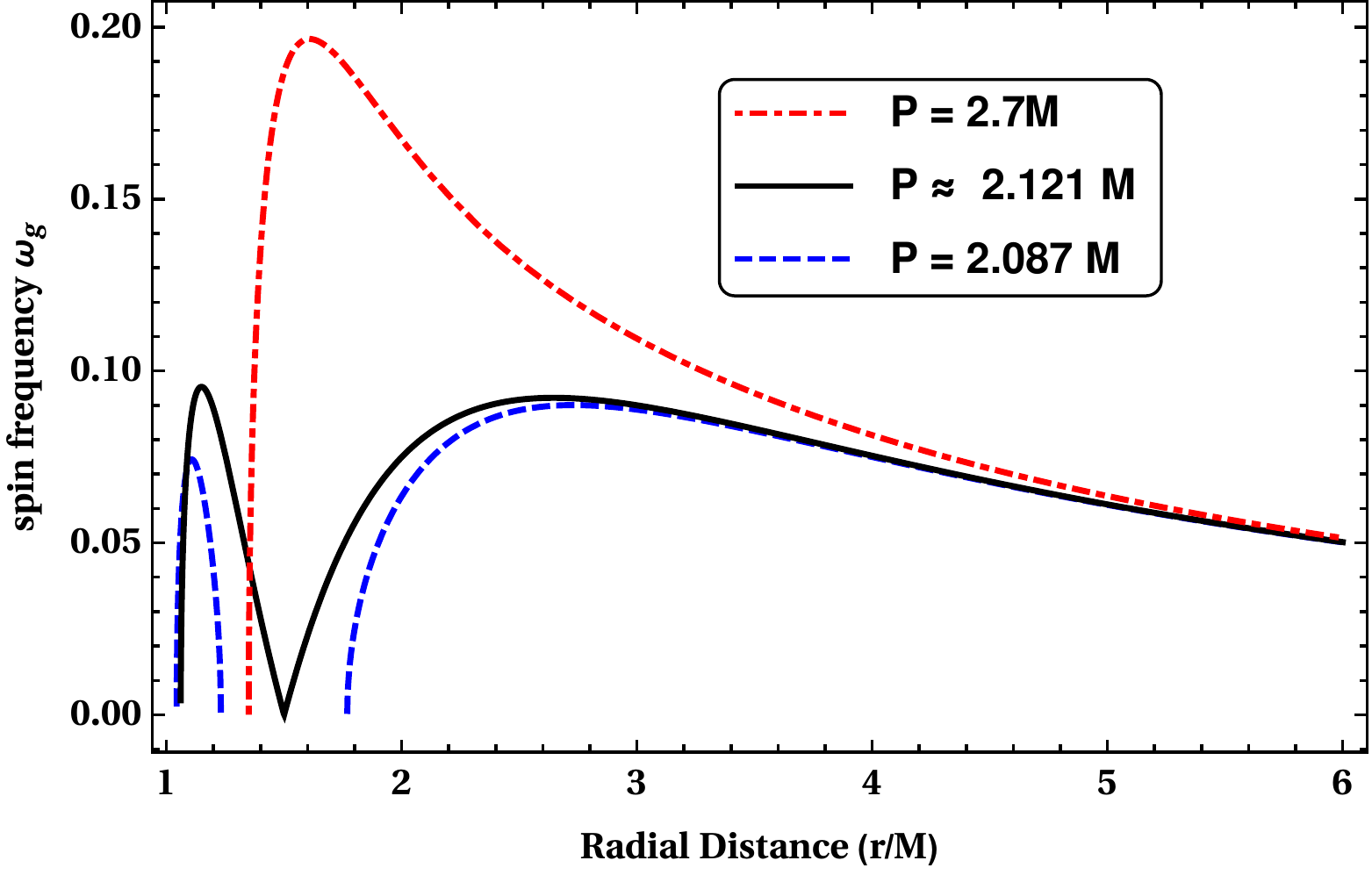}}
\caption{Geodetic spin frequency is being illustrated in the context of Sotiriou-Zhau solution.}
\label{Fig_10}
\end{figure}

Given the metric elements one can immediately compute the angular velocity of the spinning gyroscope on a circular geodesic leading to,
\begin{equation}
\omega_{\rm g}=\Omega_g\sqrt{1-\dfrac{3M}{r_{c}}+\dfrac{P^2}{2r_{c}^2}}
=\sqrt{\dfrac{M}{r_{c}^3}-\dfrac{MP^2}{4r_{c}^5}}\sqrt{1-\dfrac{3M}{r_{c}}+\dfrac{P^2}{2r_{c}^2}}~.
\end{equation}
In this case as well, the spin frequency vanishes at three locations --- (a) the two photon circular orbits ($r_{\rm ph}^{\pm}$) located at the solutions of the algebraic equation: $2r^{2}-6Mr+P^{2}=0$ as well as at (b) $r_{0}=|P|/2$, where $\Omega _{\rm g}$ also vanishes (see \ref{Fig_09}). Thus similar to the previous solution, in this case as well there can be three situations depending upon the value of the scalar charge $P$. These are:
\begin{itemize}

\item The first situation corresponds to $0<(P/M)<\sqrt{2}$. In this case both the event horizons and the photon orbits exist. Since circular geodesics are not possible within the event horizon, the gyroscope can exist only upto $r_{\rm ph}^{+}$, where the spin frequency $\omega _{\rm g}$ vanishes. This is illustrated in \ref{Fig_10a}.

\item Another possibility is to have $\sqrt{2}<(P/M)<3/\sqrt{2}$. In this case a naked singularity forms resulting into disappearance of the event horizon. However the circular photon orbit still exists. In this case the spin frequency $\omega _{\rm g}$ vanishes at three places, the outer photon orbit, the radius $r_{0}$ and at the inner photon orbit. Again the radius $r_{0}$ is not an observable, since there can be no circular geodesic in between the region $r_{0}<r_{c}<r_{\rm ph}^{+}$. This is illustrated in \ref{Fig_10b}.

\item The last possibility corresponds to $(P/M)>3/\sqrt{2}$. In this case neither the event horizon nor the photon orbit exists. Hence the spin frequency $\omega _{\rm g}$ will vanish at $r_{0}$ alone. As a consequence the radius $r_{0}$ will become an observable. Hence by just checking whether the spin frequency of a gyroscope vanishes at some radius, one may infer about the presence of a naked singularity besides the existence of scalar hair. This situation is depicted in \ref{Fig_10b}.

\end{itemize}
Having derived the spin frequency it is straightforward to compute the geodetic precession having the following expression
\begin{equation}
\mathcal{G}_{\rm g}=2 \pi \left(1-\sqrt{1-\dfrac{3M}{r_{c}}+\dfrac{P^2}{2r_{c}^2}}\right)~.
\end{equation}
The geodetic precession frequency $\mathcal{G}_{\rm g}$ of a gyroscope moving in a circular orbit takes non-trivial values, except for the photon circular orbits, where $\omega _{\rm g}$ vanishes. Since the spacetime is asymptotically flat, the geodetic precession becomes arbitrarily small at large distances, as expected. Further the geodetic precession for a gyroscope in Sotiriou-Zhou spacetime is less than the expression for \gr\ as one can easily verify. 

\begin{figure}[htp]
\subfloat[The non-geodetic frequency $\omega_{\rm nongd}$ is being plotted against radial distance for different $P$ values while $\epsilon$ is kept fixed at $0.7$. Since $\omega _{\rm nongd}$ vanishes only on the photon circular orbit, the curves for $\omega _{\rm nongd}$ will hit zero only once.\label{Fig_11a}]{\includegraphics[height=5cm,width=.49\linewidth]{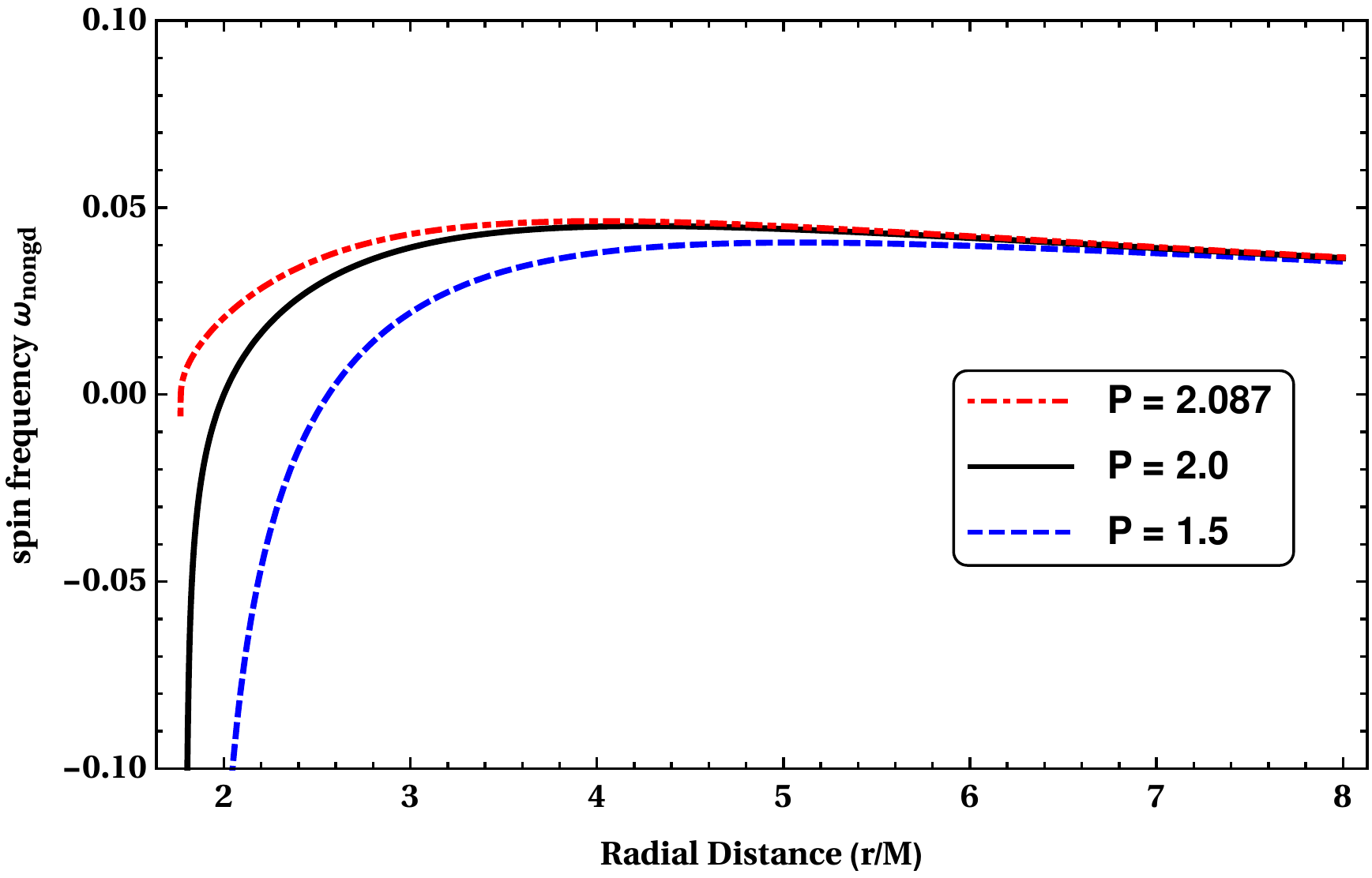}}
\hfill
\subfloat[Variation of $\omega_{\rm nongd}$ with radial distance for different choices of  $\epsilon$ is being shown, while $P$ is kept fixed at $2M$. Alike the charged Galileon black hole, the nature of the plot remains similar with an overall sign change taking place as $\epsilon$ crosses $0.5$.\label{Fig_11b}]{\includegraphics[height=5cm,width=.49\linewidth]{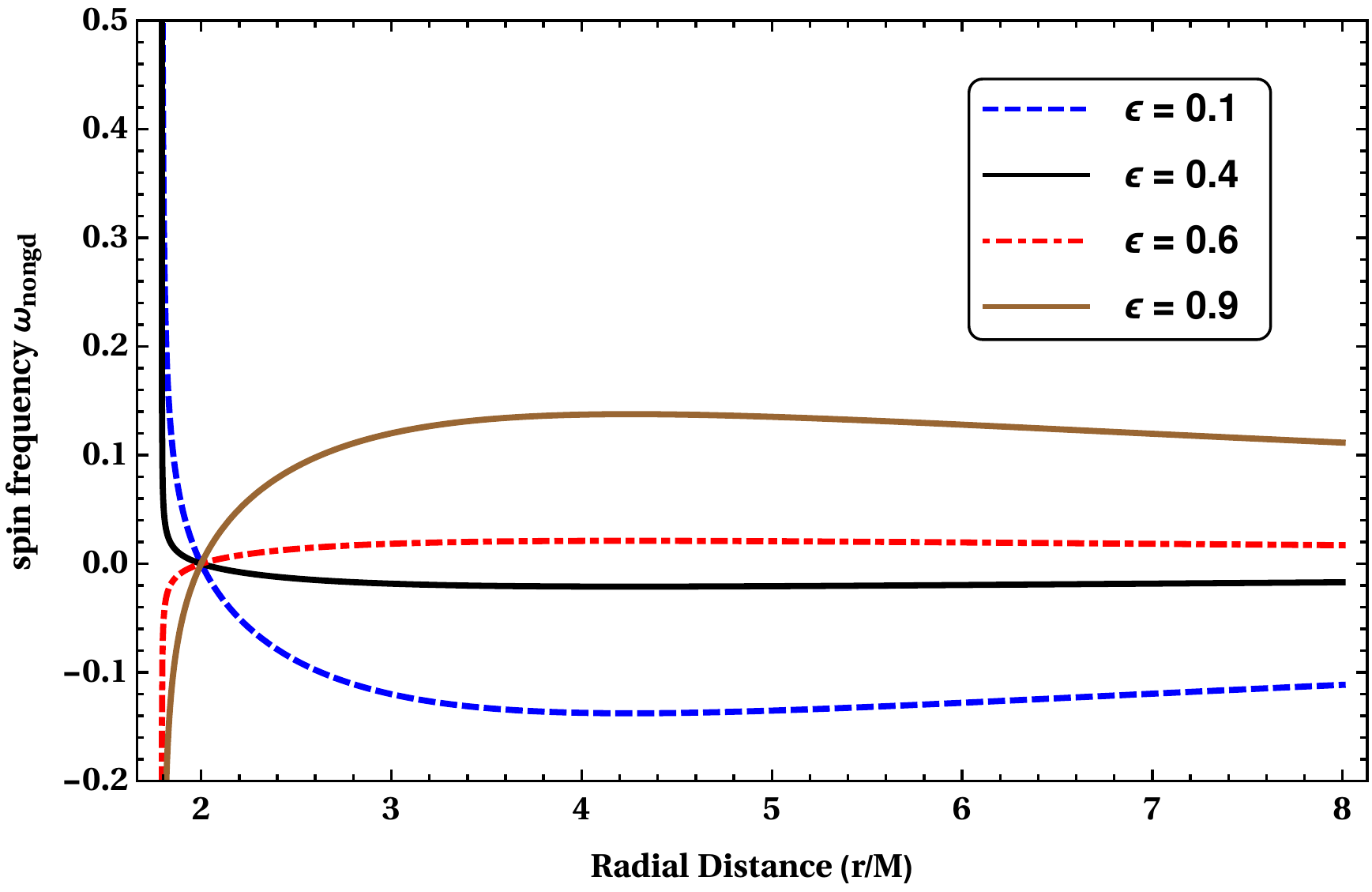}}
\\
\subfloat[The above figure illustrates the spin frequency $\omega _{\rm nongd}$ as a function of the radial distance $r_c$ and the scalar charge $P$ for non-geodesic observers with $\epsilon=0.3$. The contour representing $\omega _{\rm nongd}=0$ has also been depicted. \label{Fig_11c}]{\includegraphics[height=8.5cm,width=.49\linewidth]{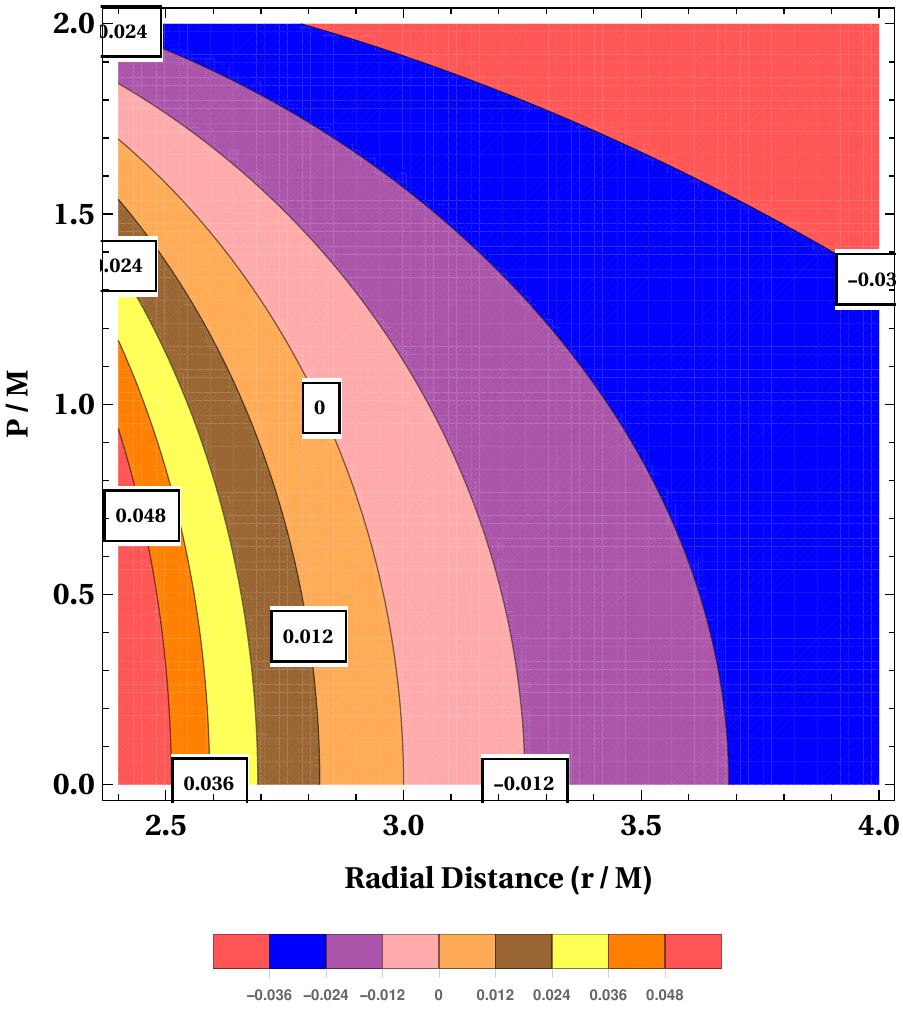}}
\hfill
\subfloat[The above figure shows the variation of spin frequency $\omega _{\rm nongd}$ with the radial distance $r_c$ and the charge parameter $q$ for non-geodesic observers with $\epsilon=0.7$. The contour with $\omega _{\rm nongd}=0$ has also been presented.\label{Fig_11d}]{\includegraphics[height=8.5cm,width=.49\linewidth]{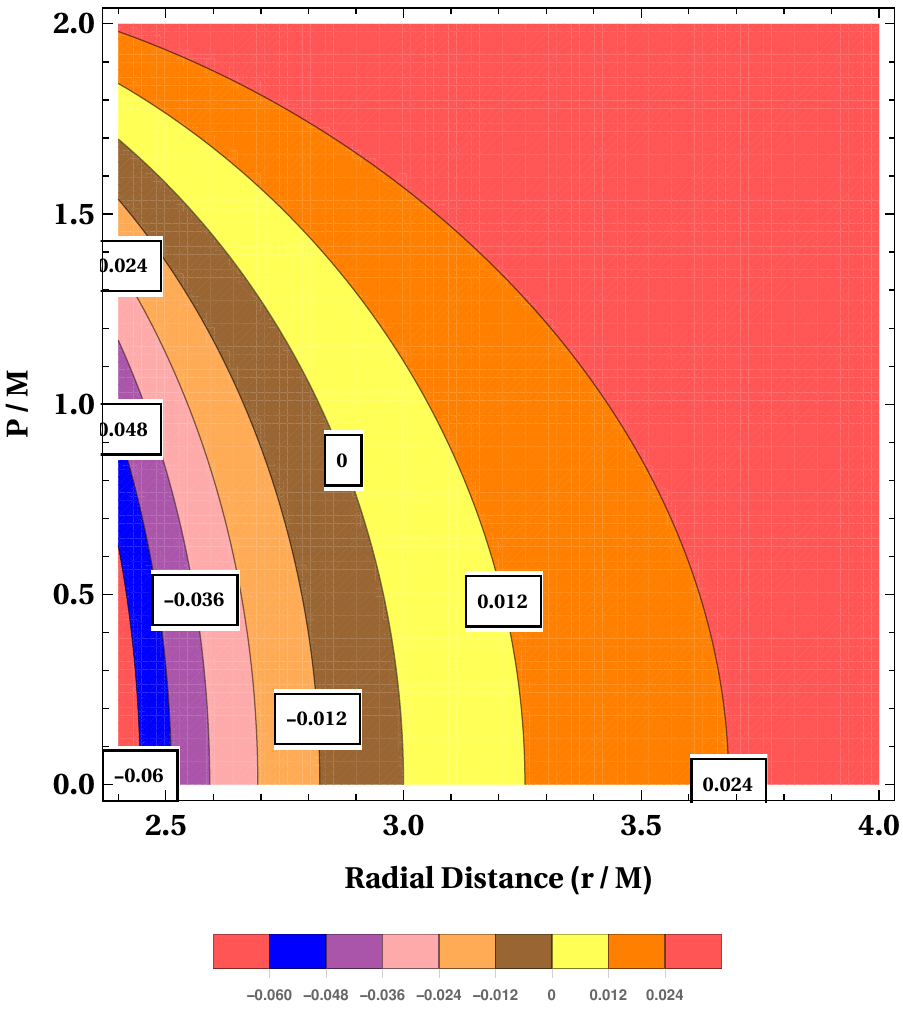}}
\caption{Spin frequency for non-geodesic observers is being shown for Sotiriou-Zhau spacetime.}
\label{Fig_11}
\end{figure}

One can smoothly carry over the analysis to non-geodesic trajectories as well. The essential steps of the computation follows the general derivation in \ref{Precession_NonGeodesic}. Using the metric components presented earlier, we arrive at the following expression for spin frequency of non-geodesic observers, using \ref{Eq:Fermi_dragged} as,
\begin{eqnarray}
\frac{\omega_{\rm nongd}}{\Omega_{\rm nongd}} &= &\dfrac{\left\{1-(3M/r_{c})+(P^2/2r_{c}^2)\right\}
\Big\{1-(2M/r_{c})+(MP^2/6r_{c}^3)\Big\}^{-1}}{\sqrt{4\epsilon\left(1-\epsilon\right)
\Big[1+(2M/r_{c})+\{(8M^2-P^2)/2r_{c}^2\}\Big]}} \\
\text{or}, \qquad \omega_{\rm nongd} &=& \dfrac{2\epsilon-1}{r\sqrt{4\epsilon\left(1-\epsilon\right)}}\dfrac{\left\{1-(3M/r_{c})+(P^2/2r_{c}^2)\right\}
\Big\{1-(2M/r_{c})+(MP^2/6r_{c}^3)\Big\}^{-1/2}}{{\Big\{1+(2M/r_{c})+[(8M^2-P^2)/2r_{c}^2]\Big \}}^{1/2}}
\end{eqnarray}
As evident from the above expression the spin frequency vanishes on the circular photon orbit but remains non-zero otherwise (see \ref{Fig_11}). Thus intriguingly the spin frequency for non-geodesic observers do not vanish anywhere when $P/M>3/\sqrt{2}$. This is a distinctive signature of Sotiriou-Zhau spacetime, essentially originating from the presence of scalar hair.
Finally use of \ref{nongeod_prec} yields the precession frequency associated with the gyroscope moving in a circular but non-geodesic trajectory as,
\begin{equation}
\mathcal{G}_{\rm nongd}=2 \pi \left(1-\dfrac{(1-3 M/r_{c}+\dfrac{P^2}{2 r_{c}^2})\left({1-2 M/r_{c}+\dfrac{M P^2}{6r_{c}^3}}\right)^{-1}}{\sqrt{1+ \dfrac{2M}{r_{c}}+\dfrac{(8M^2-P^2)}{2r_{c}^2}}\sqrt{1-(1-2 \epsilon)^2}}\right)
\end{equation}
This is also smaller compared to the corresponding expression in Schwarzschild spacetime. Note that the precession frequency is non-trivial except for the photon orbits, which is expected as $\omega _{\rm nongd}$ vanishes there. 

However, there is one issue of applying the above result pertaining to the \GPB\ experiment directly to the Sotiriou-Zhau solution. Since this particular model of Einstein-dilaton-Gauss-Bonnet gravity does not admit any star (or for that matter any perfect fluid) solution, as elaborated in the introduction. Nevertheless it opens up a very interesting avenue of exploration. Recently, there have been several observational evidences of a supermassive black hole located at the centre of the Milky Way galaxy, named Sgr A* \cite{Schodel:2002vg,Hees:2017aal}. There are several stars (in particular S2 and S6) orbiting this supermassive black hole, which can provide an ideal test bed for these alternative theories. Since these stars have intrinsic spins and they are moving in geodesic orbits around Sgr A*, the analysis presented above will become directly applicable to that situation. With the Event horizon telescope or square kilometer array becoming functional in the near future one can possibly measure the spin 
precession with better accuracy and hence will be able to constrain the respective theories much better \cite{Loeb:2013lfa,Ricarte:2014nca}. In these contexts the results derived in this work will be of considerable interest.  

As a crude estimate, if one blindly applies the results associated with the \GPB\ experiment in the context of Sotiriou-Zhau solution, then the following bound is being obtained: $(P/GM)<0.11$. Here we have re-introduced the Newton's constant. The above bound is completely consistent with the results derived in \cite{Bhattacharya:2016naa} and is within $\sim 10\%$ of the bound obtained from both perihelion precession and bending angle of light. Further note that the above scenario is directly applicable to a few more situations as well. One such scenario corresponds to black hole in the presence of \KR field \cite{Kalb:1974yc}. This also provides a hairy black hole solution identical in structure to the Sotiriou-Zhou solution and hence the above analysis will be directly applicable in this case. Thus the above constraint on scalar charge $P$ translates into the \KR field charge in that context, thereby providing yet another application of our result in a different setup. 
\section{Discussion}

The properties of a spinning gyroscope have been discussed in Horndeski theories involving arbitrary couplings between scalar and gravity, while yielding second order field equations. The fact that the equations of motion are of second order ensures that the theory is free from any Ostrogradsky ghosts, which is very much desirable \cite{Woodard:2015zca}. In this work we have explored the possibility of Horndeski theories becoming viable alternative to \gr\ in the light of geodetic precession of a spinning gyroscope and the \GPB\ experiment. Moreover as suggested earlier in \cite{Chakraborty:2016ipk,Chakraborty:2016mhx,Pradhan:2016qxa}, gyroscope can also be used as a useful probe to understand the basic structure of spacetime geometry, in particular existence of naked singularity may be inferred using spinning particles. In this work we have explored both the types of spin precession, firstly we have elaborated the motion of a spinning gyroscope along a geodesic orbit, while in the second, 
we consider a Fermi transported gyroscope orbiting in a non geodesic trajectory for a general static and spherically symmetric spacetime. The first case has been studied in detail in the context of Schwarzschild spacetime and is further supported with some experimental proofs such as \GPB, while the second one have not received much attention until Iyer and Vishveshwara \cite{Iyer:1993qa} came up with the Frenet-Serret formalism. We have employed this particular framework to understand the properties of Fermi dragged gyroscopes in the Horndeski theories.

Having developed the above formalism for a general static and spherically symmetric spacetime, we have applied the same to the Schwarzschild de-Sitter solution and have investigated the properties of a spinning gyroscope. Unlike Einstein's gravity, the features distinctly depends on the cosmological constant and shows contrasting behaviour when compared to the Schwarzschild black hole. As is well known in the case of Schwarzschild solution, the spin frequency vanishes at the photon orbit located at $r=3M$ which in fact, is closely related to the reversal of the centrifugal force \cite{abramowicz1990centrifugal,Prasanna:1990nq,abramowicz1990centrifugal2}. But, when a non-zero cosmological constant is present, the spin frequency of the gyroscope vanishes at two points, one is the usual photon orbit at $r=3M$ and another is at $r_0=(3M/\Lambda)^{1/3}$. This can be used as a probe to distinguish the de-Sitter spacetime from the asymptotically flat solutions of Einstein's gravity. In case of the Fermi dragged 
gyroscope, the precession only vanishes at the photon circular orbit and $r_{0}$ ceases to exist.

The second example discusses another exact solution of Horndeski theories, corresponding to the asymptotically flat branch of a charged Galileon black hole. Unlike the previous case, this solution is associated with non-minimal coupling of the Galileon field with gravity and a gauge field. The properties of a spinning gyroscope in this spacetime are further categorized for positive and negative values of the Galileon charge $q$. It is shown that for $q>0$, the spin frequency vanishes at the photon circular orbit along with at $r_0=qM$. But $r_0$ always remains within the outer photon circular orbit and is only visible when the photon circular orbits cease to exist. Thus when naked singularity is present, the spin frequency may vanish and one may use this fact to distinguish the existence of event horizon from naked singularity. On the other hand, in the case of $q<0$, the spin frequency of a gyroscope can only vanish at the outer photon orbit. We have also produced an useful upper bound on the parameter $q$ 
within which it obeys the findings of \GPB\ and is consistent with the previous literatures. Similar considerations apply for the asymptotically de-Sitter branch of the charged Galileon black hole as well. 

Finally we have explored the Sotiriou-Zhau solution in the context of hairy black holes in scalar coupled Einstein-Gauss-Bonnet gravity. The geometry is sharply different from the previous cases as here $g_{tt}\neq -g_{rr}$. Similar to the charged Galileon black hole, the spin frequency of the gyroscope vanishes at the outer photon orbit and as well as at $r_{0}=|P|/2$. For a large value of the scalar charge parameter $P$, when the photon circular orbit no more exists, the radius $r_{0}=|P|/2$ appears in the spacetime structure. Thus in this case as well one can differentiate between a spacetime inheriting event horizon and naked singularity by inspecting whether the spin frequency of a spinning gyroscope vanishes or not. We have also presented possible observational avenues to explore, in view of the supermassive black hole Sgr A* in the Milky Way. In particular measuring the precession frequency of the stars orbiting the supermassive black hole may provide another strong field test of 
gravity and it will be 
possible to provide more stringent constraints on the model parameters, which will either constrain them significantly or will rule them out. 

Note that our analysis have been based on spherically symmetric configuration, while a similar approach for the stationary or axi-symmetric black holes, can be obtained by a straightforward extension of the method presented here. This would be more relevant from the astrophysical point of view as black holes are likely to have angular momentum. This we leave for the future. 
\section{Acknowledgement}
One of us (S.M.) is thankful to Dr. R.K. Nayak for useful discussion on numerous occasions. He is also thankful to IACS, Kolkata for welcoming stays during his short visits. S.M. extends his gratitude to Dr. Chandrachur Chakraborty for a helpful discussion through email exchanges. Finally, research of S.C. is supported by the SERB-NPDF grant (No. PDF/2016/001589) from DST, Government of India.

\bibliographystyle{utphys1}
\bibliography{References}
\end{document}